\documentclass[preprint2, times]{aastex62}

\usepackage{graphicx}
\usepackage{natbib}
\usepackage{color}
\usepackage{epstopdf}

\def\deg {{$^{\circ}$}}
\def\wn {{cm$^{-1}$}~}
\def\wnp {{cm$^{-1}$}}
\def\um {{$\mu$m}}

\def\twonutwo {{2$\nu$$_{2}$}}
\def\nuthree {{$\nu$$_{3}$}}

\def\ph {{PH$_{3}$}}
\def\germane {{GeH$_{4}$}}
\def\gefour {{$^{74}$GeH$_{4}$}}
\def\gesix {{$^{76}$GeH$_{4}$}}
\def\gethree {{$^{73}$GeH$_{4}$}}
\def\getwo {{$^{72}$GeH$_{4}$}}
\def\gezero {{$^{70}$GeH$_{4}$}}

\def\ammonia {{NH$_{3}$}}
\def\water {{H$_{2}$O}}
\def\hs {{H$_{2}$S}}
\def\dm {{CH$_{3}$D}}
\def\methane {{CH$_{4}$}}
\def\amhs {{NH$_{4}$SH}}
\def\h {{H$_{2}$}}
\def\arcsec {{$^{\prime \prime}$}}

\shorttitle{Great Red Spot Cloud Structure}
\shortauthors{Bjoraker et al.}

\begin{document}

Received 2018 March 22; revised 2018 June 29; accepted 2018 July 2

\title{The Gas Composition and Deep Cloud Structure of Jupiter's Great Red Spot}


\author{G. L. Bjoraker}
\affil{NASA/GSFC Code 693, Greenbelt, MD 20771, USA}
\email{gordon.l.bjoraker@nasa.gov}

\author{M. H. Wong}
\affil{Department of Astronomy, University of California, Berkeley, CA 94720-3411, USA}

\author{I. de Pater}
\affil{Department of Astronomy, University of California, Berkeley, CA 94720-3411, USA}

\author{T. Hewagama}
\affil{University of Maryland, College Park MD 20742-4111, USA}

\author{M. \'{A}d\'{a}mkovics}
\affil{Department of Physics and Astronomy, Clemson University, Clemson, SC 29634-0978, USA}

\author{G. S. Orton}
\affil{Jet Propulsion Laboratory and California Institute of Technology, Pasadena, CA 91109, USA}



\begin{abstract}

We have obtained high-resolution spectra of Jupiter's Great Red Spot (GRS) between 4.6 and 5.4~\um~ using telescopes on Mauna Kea in order to derive gas abundances and to constrain its cloud structure between 0.5 and 5~bars.  We used line profiles of deuterated methane (\dm)~ at 4.66~\um~ to infer the presence of an opaque cloud \textbf{at 5$\pm1$~bars}. From thermochemical models this \textbf{is almost certainly} a water cloud. We also used the strength of Fraunhofer lines in the GRS to obtain the ratio of reflected sunlight to thermal emission. The level of the reflecting layer was constrained to be at \textbf{570$\pm30$~mbars} based on fitting strong \ammonia~ lines at 5.32~\um. We identify this layer as an ammonia cloud based on the temperature where gaseous \ammonia~ condenses. We found evidence for a strongly absorbing, but not totally opaque, cloud layer at pressures deeper than 1.3 bars by combining Cassini/CIRS spectra of the GRS at 7.18~\um~ with ground-based spectra at 5~\um. This is consistent with the predicted level of an \amhs~ cloud.  We also constrained the vertical profile of \water~ and \ammonia. The GRS spectrum is matched by a saturated \water~ profile above an opaque water cloud at 5~bars. The pressure of the water cloud constrains Jupiter's O/H ratio to be at least 1.1 times solar. The \ammonia~ mole fraction is \textbf{200$\pm50$~ppm} for pressures between 0.7 and 5~bars. Its abundance is \textbf{40~ppm} at the estimated pressure of the reflecting layer. We obtained \textbf{0.8$\pm0.2$~ppm} for \ph, a factor of 2 higher than in the warm collar surrounding the GRS. We detected all 5 naturally occurring isotopes of germanium in \germane~ in the Great Red Spot. We obtained an average value of \textbf{0.35$\pm0.05$~ppb} for \germane. Finally, we measured \textbf{0.8$\pm0.2$~ppb} for CO in the deep atmosphere.
\end{abstract}





\section{INTRODUCTION}

The Great Red Spot is a high-pressure region in the atmosphere of Jupiter, producing an anticyclonic storm 22\deg~ south of the planet's equator. The spot is large enough to contain two or three planets the size of Earth. There have been numerous studies of various aspects of the Great Red Spot (GRS) including its wind field (\citet{simon02}, \citet{asay09}), its mysterious red color (\citet{loeffler16},  \citet{carlson16}), its significant shrinkage in size (\citet{asay09}, \citet{simon14}) and its overall dynamics (e.g. \citet{marcus93}, \citet{read06}, \citet{palotai14}).

There have also been numerous studies of the vertical cloud structure of the GRS. \citet{banfield98} used Galileo Solid State Imaging (SSI) data between 0.73 and 0.89~\um~ to infer a thick cloud over the GRS extending from 200 to 700~mbars. \citet{irwin99} reported a similar cloud structure for the GRS based on \citet{baines96}. These authors used Galileo Near Infrared Mapping Spectrometer (NIMS) spectra of the GRS between 0.89 and 2.2~\um. This cloud forms at the level where \ammonia~ ice is expected to condense. However, \citet{baines02} found that spectrally identifiable ammonia clouds are present to the northwest of the GRS, but not in the spot itself. Clouds to the northwest were newly condensed and therefore fresh, while in the GRS cloud particles \textbf{may have evolved} through coating or compositional mixing (e.g. \citet{atreya05, kalogerakis08, sromovsky10}). \citet{simon01, simon02} used Galileo SSI data between 0.41 and 0.89~\um~ to infer the presence of an optically thin stratospheric haze, a moderate to dense tropospheric haze, and an optically thick, physically thin cloud sheet around 900~mbars.

Prior studies of gas composition in the GRS have mostly been limited to the upper levels of the feature, between about 100 - 500~mbar, due to opacity from aerosols and \textbf{Rayleigh} scattering. Retrievals of locally-depleted NH$_3$ in the 100 - 300~mbar range from Voyager IRIS \citep{griffith92} and HST/FOS \citep{edgington99} might seem to be at odds with
locally-enhanced NH$_3$ abundance near the 500-mbar level from
Cassini/CIRS \citep{achterberg06}. These results might be
consistent if NH$_3$ decreases rapidly with height (0.5-km scale height), as found by \citet{tokunaga80} based on ground-based infrared spectroscopy. \citet{fletcher10} used multiple sources of thermal infrared data to show that there are also
horizontal gradients of composition (as well as temperature and aerosol parameters) across the GRS. \citet{fletcher16} revisited the GRS using ground-based infrared spectral imaging with TEXES, confirming the north-south gradient of NH$_3$
concentration in the 500-mbar region. At radio frequencies, the Very Large Array (VLA) has been used to produce spectral maps of NH$_3$ concentrations at deeper levels on Jupiter, including the GRS \citep{depater16}. Wavelengths from 1.7 - 5.5~cm probe the 0.6 - 6 bar pressure levels, where the GRS has higher NH$_3$ concentrations than most other regions of the planet (except for the Equatorial Zone and plumes of NH$_3$-rich gas near 4$^{\circ}$N). At these wavelengths, gaseous NH$_3$ is the principal opacity source, while cloud opacity is expected to be minimal.

In this paper we present ground-based observations of Jupiter's Great Red Spot between 4.6 and 5.4 \um. The 5-\um~ region is a window to the deep atmosphere of Jupiter because of a minimum in opacity due to \h~ and \methane. This spectrum provides a wealth of information about the gas composition and cloud structure of the troposphere. Jupiter's 5-\um~spectrum is a mixture of scattered sunlight and thermal emission that varies significantly between Hot Spots and low-flux regions such as the Great Red Spot. Chemical models of Jupiter's cloud structure predict three distinct layers: an \ammonia~ice cloud near \textbf{0.8~bars}, an \amhs~cloud formed from a reaction of \ammonia~and \hs~at \textbf{2.3 bars}, and a massive water ice/liquid solution cloud near \textbf{6~bars}, depending on assumptions of composition and thermal structure (see \citet{weidenschilling73}, \citet{atreya85}, and \citet{wong15}). Thermal emission from the deep atmosphere is attenuated by the variable opacity at 5~\um~ of one or more of these three cloud layers.

The Great Red Spot has very low flux at 5~\um~ compared with adjacent regions. Previous 5-\um~ datasets such as Voyager IRIS, Galileo NIMS, and spectra from the Kuiper Airborne Observatory of Jupiter provided a wealth of information on Hot Spots (e.g. \citet{bjoraker86b, bjoraker86a} and \citet{roos04})  but they did not have the combination of sensitivity, high spectral resolution, and spatial resolution required to model the Great Red Spot. In this study, the use of high resolution instrumentation on the Infrared Telescope Facility (IRTF) and Keck telescopes in Hawaii allows us for the first time to characterize the cloud structure and gas composition of the Great Red Spot at pressures between 0.5 and 5~bars.

The abundance of \water~in Jupiter's atmosphere is of fundamental importance in understanding the origin of Jupiter, the composition of its clouds, and jovian dynamics at pressures greater than 2~bars. One of the key objectives of the Juno mission, which began orbiting Jupiter in July 2016, is to measure water vapor below Jupiter's clouds to determine the O/H ratio using the Microwave Radiometer (MWR) \citep{janssen05}. Interpretation of these data may not be straightforward, however, due to the small microwave absorptivity of \water~ gas compared with \ammonia~ (see \citet{depater05} for details). Ground-based measurements of  \water~ and \ammonia~ are important in order to provide upper boundary conditions to these key absorbers. Well-constrained values for these gases between 0.5 and 5~bars should improve the accuracy of the MWR's measurements of \water~ and \ammonia~ down to 100 bars, which only Juno can  perform.

In Fig.~1 we illustrate the complementarity of microwave and 5-\um~ spectra of Jupiter to probe the deep atmosphere. On the left, we show contribution functions \textbf{between 4.66~\um \ and 5.32~\um} \ due to gas opacity alone. The strongest absorption lines sound near 1 bar, while weak features can probe down to \textbf{6 or 7~bars}. Three clouds are shown at their predicted levels from thermochemical models (e.g. \citet{wong15}). On the right we illustrate weighting functions for each of the six channels of the Microwave Radiometer, adapted from \citet{janssen05}. There is excellent overlap in sounding Jupiter between 5~\um~ and 3 of the 6 MWR channels, namely at 3.125~cm, 6.25~cm, and 12.5~cm. Early Juno results on measurements of \ammonia~ in the deep atmosphere using MWR were presented by \citet{li17} and \citet{ingersoll17}. In July, 2017 Juno's orbit passed over the Great Red Spot yielding both spectacular images and microwave observations, which are currently being analyzed \citep{li17b}.

\begin{figure*}[!ht]
\begin{center}
\begin{tabular}{ll}
\hspace{-0.12in}
	\includegraphics[width=\textwidth]{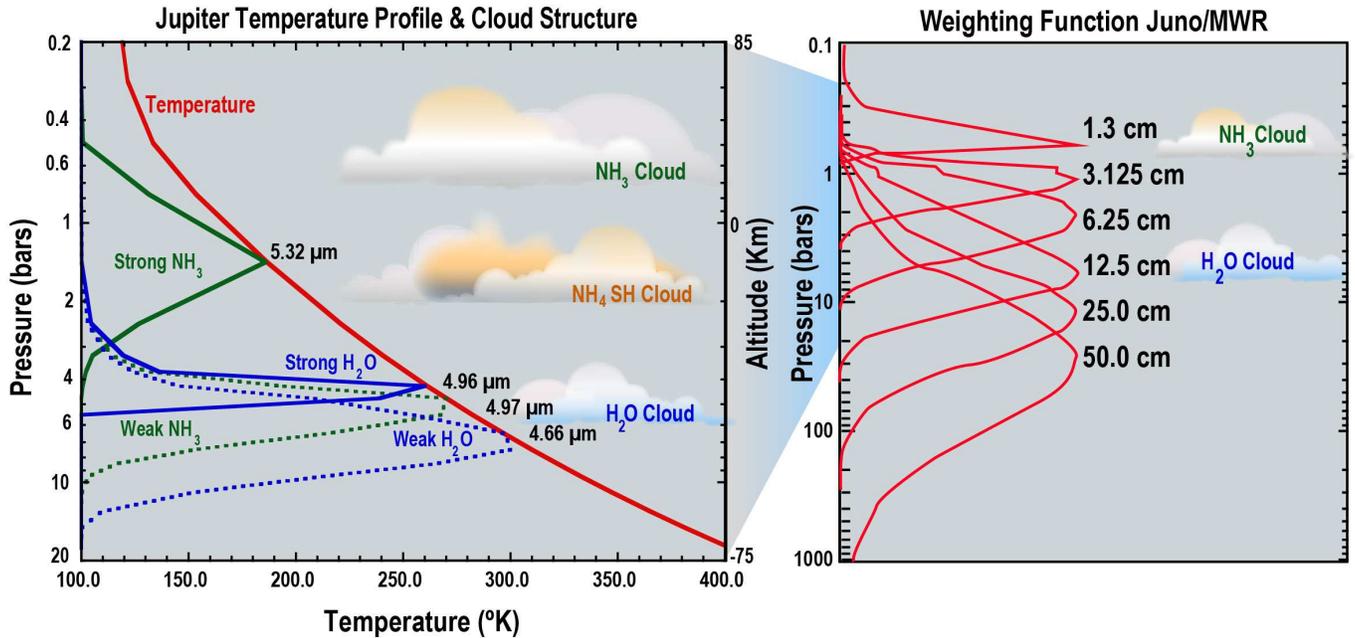}
\end{tabular}	\vspace{-0.1in}
   \caption {\footnotesize { Contribution functions of Jupiter \textbf{between 4.66~\um \ and 5.32~\um} \  compared with MWR channels on Juno \citep{janssen17}. On the left, the Galileo probe temperature/pressure profile \citep{seiff98} is shown in red for comparison.}}\vspace{-0.1in}
   \label{fig1}
\end{center}  
\end{figure*} 

In the next section we describe the instrumentation, observing circumstances, and data selection for the Great Red Spot. In Section 3 we present spectra of Jupiter's Great Red Spot from ground-based and Cassini data. We demonstrate that we can constrain the lower cloud structure of the GRS even when higher-altitude clouds greatly attenuate the thermal flux from the deep atmosphere. \citet{bjoraker15} modeled a Hot Spot in the South Equatorial Belt and a cloudy region in the South Tropical Zone. In the current study we apply the same methodology to constrain the pressure of the deepest clouds in Jupiter's Great Red Spot. In addition, we model ground-based spectra at 5.3~\um~ and Cassini spectra at 7.2~\um~ to constrain the location of upper and mid-level clouds in the GRS. Using a 3-layer highly-simplified cloud model we derive the abundance of \water, \ammonia, \ph, \germane, and CO in the deep atmosphere of the Great Red Spot. These abundances will help to understand the dynamics of the atmosphere in this unique feature on Jupiter.

\section{OBSERVATIONS}\label{obs}


Five-micron spectra of Jupiter were acquired using two instruments: iSHELL on NASA's Infrared Telescope Facility (IRTF) and NIRSPEC on the Keck 2 telescope.   iSHELL is a new echelle spectrometer with 16 orders dispersed onto a Teledyne 2048x2048 Hawaii 2RG HgCdTe array covering 4.52-5.24 \um~ using the M1 grating setting \citep{rayner16}. A 0.75\arcsec $\times$ 15\arcsec~ slit was aligned east-west on Jupiter at the latitude of the Great Red Spot, resulting in spectra with a resolving power of 35,000. We analyzed 3 orders centered on 4.66, 4.97, and 5.16~\um~ in order to retrieve cloud structure, \water, and \ammonia~ in the Great Red Spot. The iSHELL spectra were obtained on May 18, 2017. Jupiter subtended 42\arcsec~ and the geocentric Doppler shift of the center of Jupiter was 18 km/sec. By selecting a time when the GRS was near the receding limb (planeto east or sky west) the resulting GRS spectrum had a Doppler shift of 28 km/sec due to Jupiter's rapid rotation. The water vapor column above Mauna Kea was 0.68 precipitable mm derived from fitting telluric lines in both stellar and Jupiter spectra. The seeing was 0.36 arcsec.

The combination of excellent seeing and low water vapor resulted in high quality spectra of Jupiter. Fig. 2 shows an image of Jupiter at 5.1~\um~ using the slit-viewing guide camera on iSHELL. The Great Red Spot appears dark at 5~\um~ due to thick clouds. There is a warm collar surrounding the GRS where clouds are much thinner. We analyzed a portion of the collar located 3.3\arcsec~ west of the GRS. The image shows a slit that is 25\arcsec~long. However, only a 15\arcsec~ subset of the slit (denoted by vertical bars) can be used at 5~\um. Off-Jupiter spatial pixels were used to ensure proper sky subtraction. Fig. 3 is a composite image of the Great Red Spot taken with iSHELL using both the 5.1-\um~ filter and a K-band filter at 2.2~\um~ (blue). The Great Red Spot appears bright at 2.2~\um~ due to the presence of hazes in Jupiter's upper troposphere. Reflected light from deeper clouds outside of the GRS is attenuated by \methane~ and \h~ absorption. Note that excellent seeing reveals fine structure in the GRS and adjacent regions. Good seeing is also important to ensure that spectra of the GRS are not contaminated by 5-\um~ flux from adjacent regions.

\begin{figure}[!ht]
\begin{center}
\begin{tabular}{ll}
\hspace{-0.15in}
	 \includegraphics[width=3.0in]{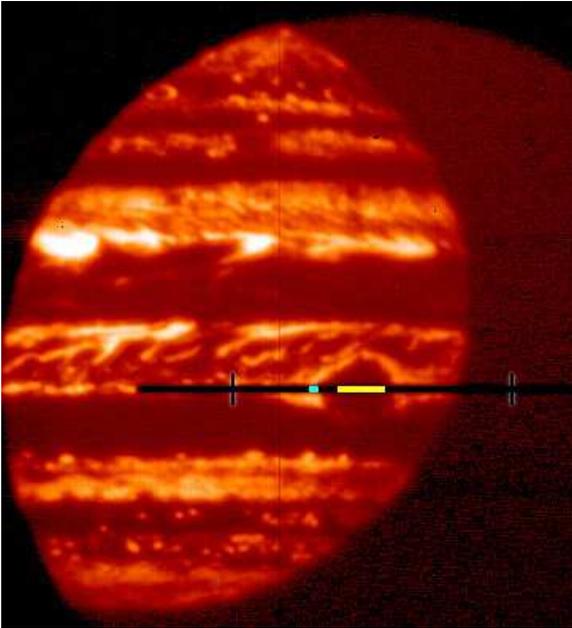}
\end{tabular}	\vspace{-0.1in}
   \caption {\footnotesize {Image of Jupiter and the GRS at 5.1~\um~ using the slit-viewing camera on iSHELL. The GRS is dark due to thick clouds that block thermal radiation. Yellow pixels denote the portion of the Great Red Spot used in this analysis. Cyan pixels show the location of the Warm Collar spectrum west of the GRS.}}\vspace{-0.2in}
   \label{fig2}
\end{center}  
\end{figure}

\begin{figure}[!ht]
\begin{center}
\begin{tabular}{ll}
\hspace{-0.15in}
	 \includegraphics[width=3.0in]{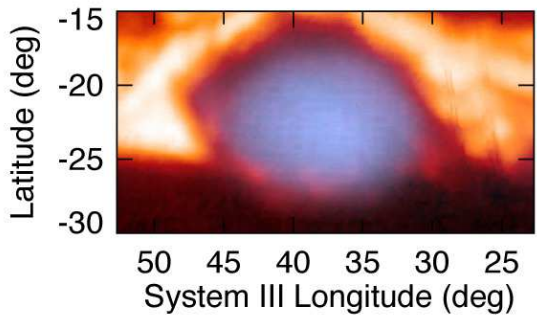}
\end{tabular}	\vspace{-0.1in}
   \caption {\footnotesize {Composite map of GRS at 2.2~\um~ (blue) and at 5.1~\um~ using the slit-viewing camera on iSHELL. The seeing was 0.36 arcsec. The GRS is bright in methane-band images due to upper tropospheric haze around 200 mbars.}}\vspace{-0.2in}
   \label{fig3}
\end{center}  
\end{figure} 

NIRSPEC is an echelle spectrograph with 3 orders dispersed onto a 1024x1024 InSb array at our selected grating/cross-disperser settings of 60.48 / 36.9 \citep{mcLean98}.  A 0.4\arcsec $\times$ 24\arcsec~ slit was aligned east-west on Jupiter at the latitude of the Great Red Spot, resulting in spectra with a resolving power of 20,000. We analyzed all 3 orders centered on 4.66, 4.97, and 5.32~\um. The first two orders yielded spectra similar to that of iSHELL. The 5.32-\um~ order samples a strong \ammonia~ absorption band permitting retrieval of the pressure of the upper cloud layer. The NIRSPEC spectra were obtained on January 21, 2013. Jupiter subtended 44\arcsec~ and the geocentric Doppler shift of the GRS was 14.3 km/sec.  The water vapor column above Mauna Kea was 1.0 precipitable mm, derived from fitting telluric lines in both stellar and Jupiter spectra.

The third dataset used in this study was a selection of thermal infrared spectra acquired during the Cassini flyby of Jupiter from December 2000 to January 2001. CIRS was the Composite Infrared Spectrometer \citep{flasar04}, operating between 6.7 and 1000~\um. We used mid-IR spectra of the Great Red Spot and of Hot Spots in the North Equatorial Belt covering wavelengths between 6.7 and 9.5~\um~(1050 to 1495 \wn). The spectral resolution was 3.0~\wn. We averaged 22 of the hottest CIRS spectra in the NEB acquired at the highest spatial resolution (2.4\deg~ of latitude) on December 31, 2000. Next, we averaged 23 of the coldest spectra of the GRS ranging between 17\deg S and 23\deg S and 41 to 61\deg~ System III longitude with an average spatial resolution of 3\deg. We include these spectra to distinguish between cloud opacity in the \ammonia~ and \amhs~ layers, as described in Section 3.1.3.

\begin{figure*}[!ht]
\begin{center}
\begin{tabular}{ll}
\hspace{-0.15in}
	 \includegraphics[width=\textwidth]{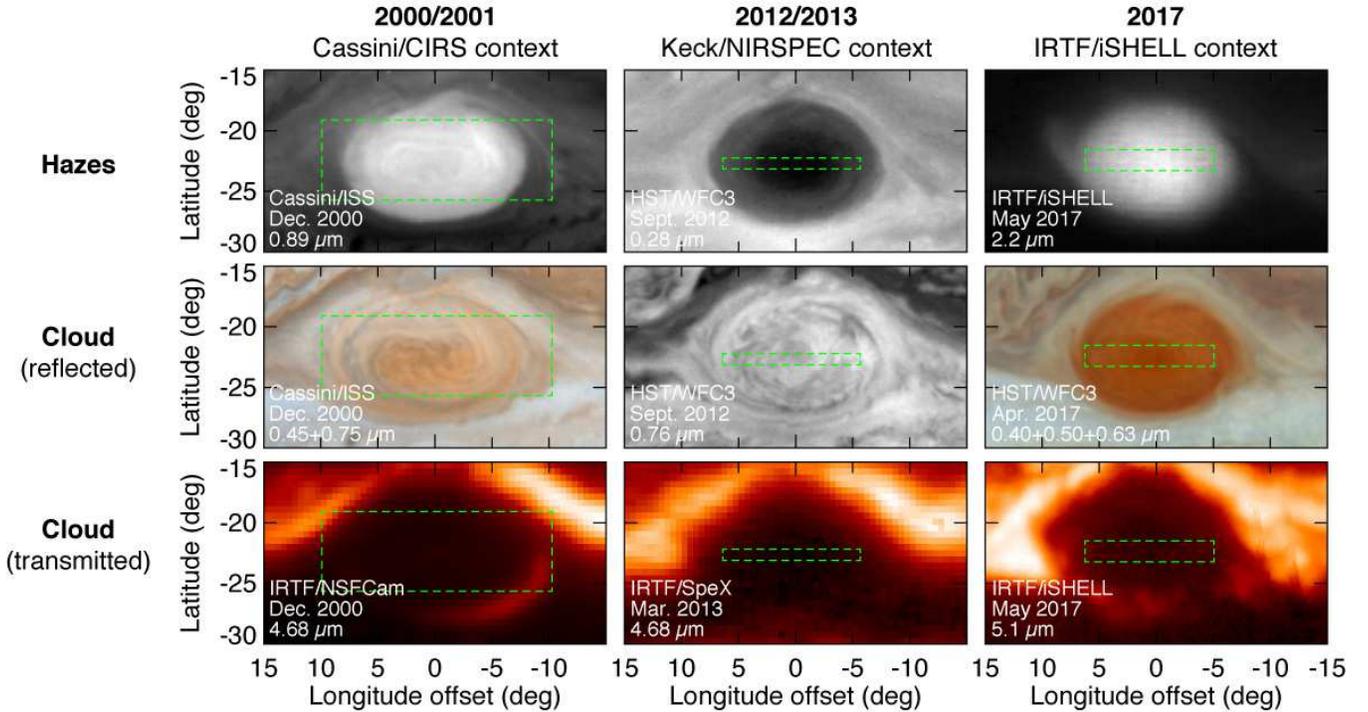}
\end{tabular} \vspace{-0.1in}
\caption  {\footnotesize {Context maps for GRS spectroscopic data. Each column corresponds to a different
epoch, and each row senses different vertical layers. Green rectangles outline
approximate areas used for spectral selection. Top row: Upper tropospheric hazes
are isolated by strong CH$_4$/H$_2$ absorption or by Rayleigh scattering in the
UV. Haze/chromophore boundaries have been shown to closely match the dynamical
boundary of the vortex \citep{depater10,wong11}. Our spectral footprints
are within these boundaries. The CIRS spectral footprint (left column) appears
in this figure to extend beyond the dynamical boundary of the vortex, but only
the coldest spectra within this footprint were selected. In the UV (middle
column), the haze within the GRS has a low single-scattering albedo, causing its
central region to appear darker than the surroundings. Middle row: Overall cloud
opacity in the NH$_3$ and NH$_4$SH layers modulates optical reflectivity. Bottom
row: M-band brightness is shown on a log scale (top/middle rows are linear
scale). Here, bright regions suggest low cloud opacity in both NH$_3$ and
NH$_4$SH layers.  Maps are shown for relative context and are not
radiometrically calibrated. Sources: Cassini/ISS maps were obtained from http://ciclops.org. HST images in 2012 are from program GO-13067 \citep{karalidi13}, and the HST composite image from 2017 is from the OPAL program GO-14756 \citep{simon15b}. IRTF maps from 2000 and 2013 are from co-author Orton. iSHELL maps for 2017 were obtained as guider images during our spectroscopic observations.}}
\vspace{-0.1in}
   \label{fig4}
\end{center}  
\end{figure*} 

In Fig.~4 we present context images of the Great Red Spot for each of the three spectroscopic datasets used in this study. The three columns correspond to the times of the Cassini flyby of Jupiter, of the Keck/NIRSPEC observations, and of the IRTF/iSHELL data, respectively. Each row sounds a different range of altitudes. Hazes in the upper troposphere near 200~mbars are bright at 0.89~\um~ and at 2.2~\um~ and dark in the ultraviolet. Chromophores mixed into the haze absorb strongly in both ultraviolet and blue wavelengths giving the GRS its red color (see \citet{wong11, loeffler16}). Ammonia and ammonium hydrosulfide clouds can be probed by reflected sunlight using filters in the optical and near-infrared. Finally, multiple cloud layers can be studied using thermal emission at 5~\um~ that originates between 5 and 10 bars and is attenuated by \water, \amhs, and \ammonia~ cloud opacity.

\section{DATA ANALYSIS AND RESULTS}
\subsection{Cloud Structure}

We present in this section a cloud model for the Great Red Spot that for the first time extends down to 5 bars on Jupiter. Studies of the GRS in the optical and in the mid-infrared are sensitive only to clouds at pressures less than 1~bar (presumably ammonia ice) and microwave measurements are not sensitive to clouds at all (unless they contain centimeter-size particles). The 5-\um \ window is the only spectral region that is sensitive to the levels of all 3 clouds (\ammonia, \amhs, and \water) predicted from thermochemical models. In order to simplify the radiative transfer calculations, we made several approximations. The lower cloud was assumed to be semi-infinite, the upper clouds were modeled as purely absorbing, rather than as scattering layers, and each cloud layer was taken to be vertically thin (particle scale height much less than the gas scale height).

\textbf{Our treatment of clouds as non-scattering absorbing/emitting layers may result in errors in the estimated transmittances of the cloud layers. Due to the strong dependence of the Planck function on temperature, the neglected scattering component should be dominated by radiation from below each cloud layer (rather than back-scattered radiation from above the clouds). Microphysical upper limits to the radius of NH$_3$ and NH$_4$SH cloud particles are roughly 10 to 30~$\mu$m \citep{carlson88}; therefore, the scattering size parameter $x$ of cloud particles ($x = 2 \pi r / \lambda$) should be less than 40 at 4.6~\um, placing these particles in the Mie scattering regime in the absence of strong shape effects \citep{mishchenko98}. The Mie single-scattering phase function is dominated by forward scattering, so both phase function and thermal profile arguments suggest that our scattering-free treatment is likely to lead to somewhat overestimated cloud transmittances. The dominance of upward-scattered emission over back-scattered emission in the spectrum means that our modeled gas abundances are not strongly affected by treatment of cloud layers. Gas abundances are derived from line shapes, and these should be preserved in the scattering process, if the scattered component is dominated by upwelling radiation.} We now discuss each of the three clouds, beginning with the deepest one.

\subsubsection{Lower cloud level}

In the absence of clouds, the spectrum of Jupiter at 5~\um~ is formed between 1 and 10~bars, as shown in Fig.~1. Weak absorption lines near 4.67~\um~ (2142~\wnp) sound levels deeper than 4~bars where temperatures are greater than 250~K. When cloud opacity is introduced to our model, the effect upon the spectrum depends on the temperature at the cloud level. Cold clouds, with temperatures less than 200~K, have the effect of attenuating the continuum without changing line to continuum ratios of weak absorption lines. This is because the radiance of a 250~K black body is 22 times that of a 200~K black body at 2142~\wnp. A purely absorbing cloud of unit optical depth at 200~K will have the effect of multiplying the transmittance of a gas-only warm model atmosphere by 0.37. Self emission from the cold cloud is negligible at 2142~\wnp. Warm clouds with temperatures greater than 250~K, on the other hand, will have significant self emission. This will affect the strength and shape of absorption lines in this spectral region. Thus, the continuum level is sensitive to the total cloud opacity in Jupiter's atmosphere while spectral line shapes are sensitive only to the deepest (warmest) clouds. In this section we investigate the effect of the lower clouds on spectral line shapes. We match the observed continuum by multiplying the calculated spectrum by a constant factor to simulate purely absorbing upper (\ammonia) and mid-level (\amhs) clouds.

The 4.7-\um~ spectrum of the GRS contains absorption lines of deuterated methane (\dm), phosphine (\ph), and \water~ in Jupiter's troposphere as well as Fraunhofer lines due to CO in the Sun. Jupiter is too warm for methane and its isotopologues to condense. Photochemical destruction of \methane~ occurs at pressures less than 1~mbar in Jupiter's stratosphere, but not in the troposphere \citep{moses05b}. We therefore assume that \methane~ and \dm~ have a constant mixing ratio with respect to \h~ in Jupiter's troposphere. In that case variations in the strength and shape of  \dm~lines between the GRS and Hot Spots on Jupiter are due only to changes in cloud structure, not gas concentration as described in \citet{bjoraker15}.

Fig. 5 shows iSHELL spectra at 4.66 \um~ (2144 \wnp) of the Great Red Spot and of the Warm Collar at the same latitude 3.3\arcsec~ to the west of the GRS. Due to excellent seeing, there is no contamination of the GRS spectrum from this or other warm regions. We performed an additional check for flux contamination by comparing the spectrum of the center 3 pixels (0.5\arcsec) of the GRS with the average of 15 pixels (2.5\arcsec) shown in Fig.~5. The spectra are nearly identical, but the signal-to-noise is better in the larger average. Note that \dm~ lines are much broader in the Warm Collar than in the GRS. \textbf{The radiance of the Warm Collar spectrum (right axis) is 13 times that of the GRS (left axis) at 2142~\wnp}. The \dm~ lines are spectrally resolved at the iSHELL resolving power of 35,000 (0.06~\wn resolution). This is because molecules such as \dm \ typically have broadening coefficients of 0.06~\wnp/atm and the line formation region on Jupiter at 4.66~\um \ takes place at pressures greater than 2 bars. In addition to \dm,  this portion of the spectrum includes numerous Fraunhofer lines (denoted by S for solar). These lines are observed only in low-flux zone regions such as the GRS and not in Hot Spots. They provide evidence for a reflected solar component to the GRS spectrum at 5~\um. Spectra of the Warm Collar and of Hot Spots, on the other hand, are dominated by thermal emission.

\begin{figure}[!ht]
\begin{center}
\begin{tabular}{ll}
\hspace{-0.15in}
	 \includegraphics[width=3.0in]{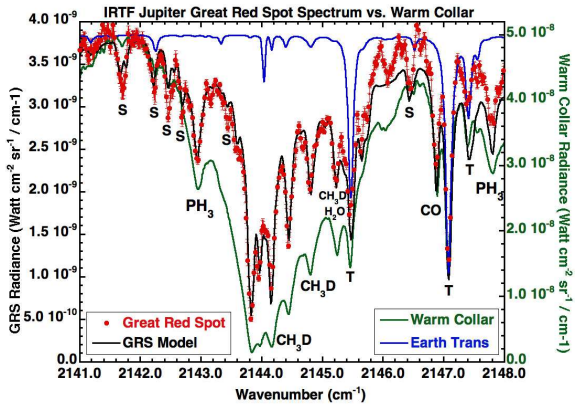}
\end{tabular}	\vspace{-0.1in}
   \caption {\footnotesize {Comparison between IRTF / iSHELL spectra of the Great Red Spot \textbf{(left axis) and of the Warm Collar (right axis) at 4.66~\um. The Warm Collar is 13 times brighter than the GRS}. Also shown is the best fitting model of the GRS and a model of the transmittance of the atmosphere above Mauna Kea. Transmittance scale (not shown) is 0 to 1.04. Telluric lines are denoted by T and solar lines by S.}}\vspace{-0.2in}
   \label{fig5}
\end{center}  
\end{figure}

\textbf{Noise at 5~\um \ is primarily due to thermal emission of the relatively warm telescope and atmosphere above Mauna Kea, rather than detector noise or source noise. In a 5-second integration, there were 8000 counts on the sky (in strong water lines) and only 100 sky-subtracted counts on the GRS. We calculated the noise in the iSHELL spectrum of the Great Red Spot in two different ways. First, we measured the standard deviation of 70 spectral points near 2141~\wn in a spatial pixel that was well off of the limb of Jupiter. Next, we measured the standard deviation of the 15 spatial pixels that were used to create an average GRS spectrum, also near the continuum frequency at 2141~\wn. We divided by the square root of 15 to determine the standard error of the mean at each frequency. The first technique resulted in a signal to noise ratio (S/N) of 51 on the continuum. The second approach yielded S/N = 53. In Fig.~5 we show error bars for each spectral frequency using the second technique}.

\begin{figure}[!ht]
\begin{center}
\begin{tabular}{ll}
\hspace{-0.15in}
	 \includegraphics[width=3.0in]{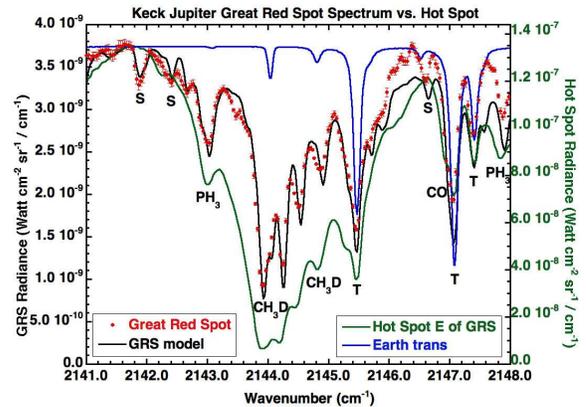}
\end{tabular}	\vspace{-0.1in}
   \caption {\footnotesize {Comparison between Keck / NIRSPEC spectra of the Great Red Spot \textbf{(left axis) and of a Hot Spot (right axis) at 4.66~\um. The Hot Spot is 36 times brighter than the GRS}. The NIRSPEC spectrum of the GRS is nearly identical to that obtained by iSHELL. The same radiative transfer model fits both Keck and IRTF spectra of the GRS. It was only necessary to change the spectral resolution, Doppler shift, and telluric water abundance to match each dataset.}}\vspace{-0.2in}
   \label{fig6}
\end{center}  
\end{figure}

Fig. 6 shows a very similar \textbf{GRS} spectrum acquired using NIRSPEC on the Keck telescope in 2013. Here we used the center \textbf{15 pixels (2.9\arcsec)} of the GRS. The similarity of the GRS spectra using two different instruments 4 years apart shows the reproducibility of the data. Also shown is the spectrum of a Hot Spot 12.7\arcsec \ east of the GRS. \textbf{As was the case with the Warm Collar, the \dm \ lines are much broader in the Hot Spot than in the GRS. The radiance of the Hot Spot spectrum (right axis) is 36 times that of the GRS (left axis) at 2141.7~\wnp}. Again, we are confident that scattered light from adjacent regions is not getting into the GRS spectrum due to good seeing (1.0\arcsec) and due to the sheer size of the GRS (see Fig.~4). \textbf{A comparison between Fig.~5 and Fig.~6 reveals more Fraunhofer lines (marked s) in the iSHELL spectrum of the GRS. These particular Fraunhofer lines are not spectrally resolved at iSHELL resolution. The factor of 2 increased resolution of iSHELL with respect to NIRSPEC makes these lines much easier to detect on Jupiter. Higher spectral resolution also helps to separate jovian CO from telluric lines and to spectrally resolve the \dm~ lines in the GRS.}

\textbf{We applied the same approach as before to determine the noise in the NIRSPEC spectrum of the GRS. We obtained a continuum S/N of 66 using off-Jupiter pixels and 80 using the variation between the 15 spatial pixels comprising the GRS spectrum. In Fig.~6 we show error bars for each spectral frequency using the second technique}.

Absolute flux calibration was performed by smoothing the NIRSPEC spectrum of an NEB Hot Spot to 4.3 \wn resolution, dividing by the transmittance of the Earth's atmosphere above Mauna Kea, and scaling the radiance of the resulting spectrum to an average of Jupiter's NEB Hot Spots observed by Voyager IRIS in 1979. We performed a general validation of this 
Voyager-scaling calibration method by checking it against a Mars-based calibration. On March 25, 2014 we observed 
an SEB Hot Spot as well as Mars at 4.66~\um~ using the predecessor to iSHELL (CSHELL). \textbf{The Mars Climate Database (MCD) is a database of meteorological fields derived from General Circulation Model (GCM) numerical simulations of the Martian atmosphere \citep{millour15}. This database can be used to calculate the surface temperature of Mars at any season. We used the web interface to MCD} ($http://www-mars.lmd.jussieu.fr/mcd-python/$) to determine that the surface temperature near the sub-solar point on Mars was 279~K at the time. Using a 279-K black body for Mars, we determined that the continuum of the SEB Hot Spot at 2142~\wn corresponded to a brightness temperature of 271.3~K. We found that the Mars calibration and our Voyager-scaling method agreed to within 10\%. The iSHELL spectrum of the GRS was normalized at 2142~\wn~ to the same radiance as in the NIRSPEC GRS spectrum.

There are numerous Fraunhofer lines visible in GRS spectra shown in Figures 5 and 6. We compared the equivalent width of the strongest line at 2141.8~\wn with its measured value in the Sun using ATMOS data \citep{farmer89, farmer94}. This line is 60\% as strong as in the Sun; thus, 40\% of the flux consists of thermal emission originating in the deep atmosphere that has been attenuated by one or more cloud layers before escaping to space.  By multiplying the radiance of the continuum of the GRS at 2142~\wn by 0.6, we obtain the value of reflected radiance from upper clouds. The albedo of the GRS can be calculated by scaling the solar flux measured at Earth to 5.07 AU (Jupiter's heliocentric distance at the time of the NIRSPEC data). The reflected radiance is proportional to the cloud albedo and inversely proportional to the square of Jupiter's heliocentric distance. The result is that the albedo of the reflecting layer over the GRS is 17\%~ at 4.66~\um.

Synthetic spectra were calculated using the Spectrum Synthesis Program (SSP) line-by-line radiative transfer code as described in \citet{kunde74}.  The input temperature profile was obtained from the Galileo Probe (Seiff et al. 1998). Line parameters for \dm~ and other 5-\um~ absorbers are from GEISA 2003 \citep {husson05}. Parameters for \dm-\h~and \dm-He broadening have been measured in the lab \citep{boussin99, lerot03, fejard03}. We used a broadening coefficient of 0.0613~\wn/atm (296/T)$^{0.5}$  for \dm~ colliding with a mixture of 86.3\%~\h~ and 13.6\%~helium, as measured by the Galileo Probe \citep{vonzahn98}. Pressure-induced \h~ coefficients were obtained using laboratory measurements at 5 \um~ by \citet{bachet83} and the formalism developed by \citet{birnbaum76}.

We investigated the effect of opaque lower clouds at 2, 5 and 7~bars on the 4.66-\um~ spectrum of the Great Red Spot. These levels correspond to opaque \amhs~ clouds and to opaque \water~clouds at two different levels. \textbf{Synthetic spectra were calculated for a gas composition of 0.18~ppm \dm \ \citep{bjoraker15, lellouch01}} and a saturated profile of \water~ above each lower cloud. The mole fraction of \ph~ was iterated to a value of 0.8~ppm to match the absorption feature at 2143 \wnp, as described in Section 3.2.3.

The calculated GRS spectrum was split into two parts. The thermal component was convolved to 0.02~\wn and Doppler-shifted by 28 and 14.3 km/sec for iSHELL and NIRSPEC data, respectively. The reflected solar component was calculated using the same model but using only transmittances above a reflecting layer at 570~mbar (see next section) and for an albedo of 17\% (as described above). The reflected component was convolved and Doppler-shifted. It was then multiplied by the ATMOS solar spectrum smoothed to 0.02~\wn resolution. The thermal component was multiplied by transmittances of 0.4, 0.0185, and 0.0134 for models with an opaque cloud at 2, 5, and 7~bars, respectively, and added to the reflected spectrum. These values represent the transmission of cold, upper clouds. This, in turn, was multiplied by the transmission of the Earth's atmosphere above Mauna Kea at the time of the iSHELL or NIRSPEC data and finally smoothed to 0.06~\wn resolution to match iSHELL or 0.10~\wn  to fit NIRSPEC spectra.

In Figures 7 and 8 we illustrate that the model with an opaque \amhs~cloud at 2 bars (gold curve) does not fit the observed spectrum. In contrast, all \dm~ features, including the wing between 2143.2 and 2143.7~\wn are equally well matched by models with opaque clouds at 5 and 7~bars. The model with a deep cloud at 7 bars fits portions of the spectrum but the model with an opaque cloud at  5~bars (black curve) provides a better fit to the spectrum near 2138~\wn, as shown in Fig.~8.  \textbf{We conclude that there must be significant cloud opacity at 5$\pm 1$~bars in the GRS. Based on the temperatures in this pressure range (257-290~K)}, thermochemical models rule out compositions of \ammonia~ or \amhs, so this cloud must consist of water in the ice or liquid phase.

\begin{figure}[!ht]
\begin{center}
\begin{tabular}{ll}
\hspace{-0.15in}
	 \includegraphics[width=3.0in]{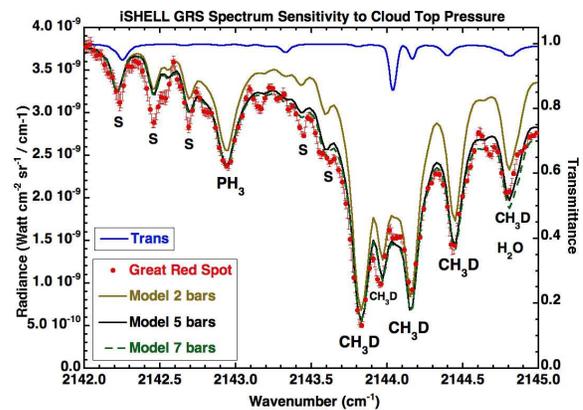}
\end{tabular}	\vspace{-0.1in}
   \caption {\footnotesize {Comparison between IRTF / iSHELL spectrum of the Great Red Spot at 4.66~\um~ and three models that differ only in the pressure of the lower boundary, assumed to be an opaque cloud. A model with an opaque cloud at 2 bars is excluded, while in this spectral region models with an opaque cloud at 5 and 7 bars fit equally well.}}\vspace{-0.2in}
   \label{fig7}
\end{center}  
\end{figure}   

\begin{figure}[!ht]
\begin{center}
\begin{tabular}{ll}
\hspace{-0.15in}
	 \includegraphics[width=3.0in]{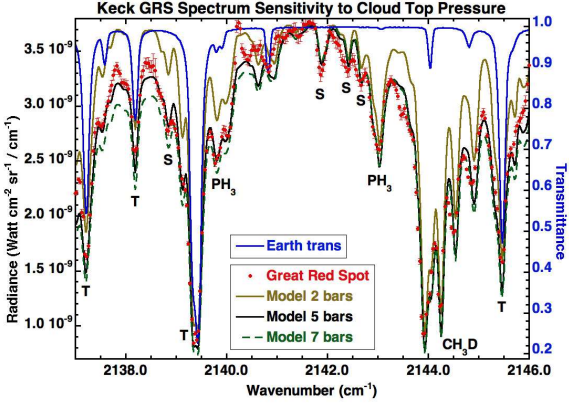}
\end{tabular}	\vspace{-0.1in}
   \caption {\footnotesize {Comparison between Keck / NIRSPEC spectrum of the Great Red Spot at 4.66~\um~ and three models that differ only in the pressure of the lower boundary. A model with an opaque cloud at 2 bars is excluded. The model with an opaque cloud at 5 bars fits the region near 2138~\wn~ better than the model with an opaque cloud at 7 bars.}}\vspace{-0.2in}
   \label{fig8}
\end{center}  
\end{figure}   

In Figure 9 we explore the sensitivity of the GRS spectrum to the fraction of reflected sunlight from the upper cloud layer. Changes in this parameter affect the strength of Fraunhofer lines as well as \dm~ absorption features. \textbf{The best fit to the 4.66-\um~ spectrum requires a mixture of 0.6$\pm0.1$ reflected sunlight and 0.4$\mp0.1$ thermal emission. This corresponds to an upper cloud albedo of 0.17$\pm0.03$}. Models with fractions of 0.4 and 0.8 reflected sunlight (corresponding to albedos of 0.115 and 0.23) do not fit the \dm~ absorption lines, nor the \ph~ feature at 2143~\wn nearly as well.

\begin{figure}[!ht]
\begin{center}
\begin{tabular}{ll}
\hspace{-0.15in}
	 \includegraphics[width=3.0in]{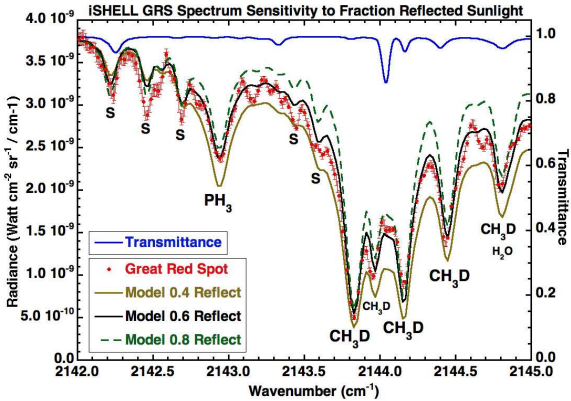}
\end{tabular}	\vspace{-0.1in}
   \caption {\footnotesize {Comparison between IRTF / iSHELL spectrum of the Great Red Spot and three models that differ  in the fraction of reflected sunlight at 4.66~\um. This parameter affects the strength of solar lines (S) as well as \dm~ features. The model with 0.4 reflected sunlight requires an upper cloud albedo of 0.115. The model with 0.8 reflected sunlight requires an upper cloud albedo of 0.23. \textbf{The best fit has an albedo of 0.17$\pm0.03$, resulting in a fraction of 0.6$\pm0.1$ reflected sunlight.} }}\vspace{-0.2in}
   \label{fig9}
\end{center}  
\end{figure} 

\subsubsection{Upper cloud level}

The 5-\um~ spectrum of Jupiter's Great Red Spot is a combination of thermal emission and reflected sunlight. To isolate the reflected component we used the very strong \twonutwo~band of gaseous \ammonia~ at 5.32~\um~ (1880~\wnp). Here, absorption by \ammonia~ is so strong that the thermal component is essentially zero. \textbf{The continuum level is set by the albedo of the reflecting level (0.17$\pm0.03$), which we measured at 4.66~\um~ using Fraunhofer lines. These lines are blended with \ammonia~ and so we cannot use their strength to derive the albedo of the GRS at 5.32~\um. We adopted an albedo of 0.17$\pm0.03$ across the entire 5-\um~ spectrum}, a reasonable assumption because the imaginary index of refraction for \ammonia~ ice is between 0.01 and 0.001 in this wavelength range \citep{martonchik84}.

\begin{figure}[!ht]
\begin{center}
\begin{tabular}{ll}
\hspace{-0.15in}
	 \includegraphics[width=3.0in]{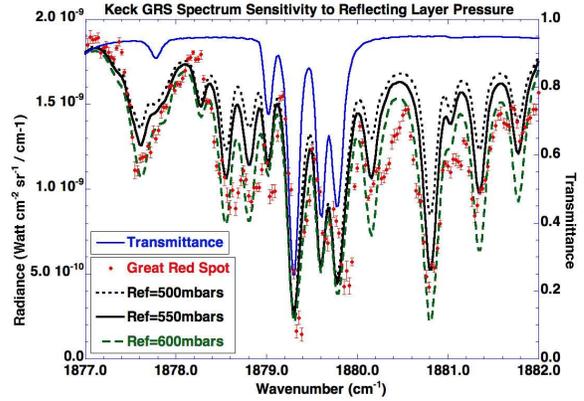}
\end{tabular}	\vspace{-0.1in}
   \caption {\footnotesize {Comparison between NIRSPEC spectrum of the Great Red Spot at 5.32~\um~ and three models with reflecting layers at 500, 550, and 600~mbars. All features are due to ammonia except for the telluric water lines shown in blue. The mole fraction of \ammonia~ is 14, 30, and 60~ppm at these levels. \textbf{The best fit is for a reflecting layer at 570$\pm 30$~mbars.}}}\vspace{-0.2in}
   \label{fig10}
\end{center}  
\end{figure}

In Fig.~10 we show the spectrum of the Great Red Spot where the strongest \ammonia~ features occur. These absorption features are sensitive to the column of \ammonia~ above the reflecting layer. However, \ammonia~ is condensing so we cannot measure the vertical profile of ammonia and the pressure of the reflecting layer separately. Therefore, we assume that the mole fraction of \ammonia~ is 200~ppm in the deep atmosphere (see Section 3.3) and follows the saturated vapor pressure relation for pressures less than 700~mbars. We used a temperature profile obtained from the Galileo Probe \citep{seiff98}. Although this profile was not measured in the GRS, there is good agreement between the Probe temperature of 130.0~K at 468~mbars and a value of 130~K at 400~mbars within the GRS derived from mid-IR mapping \citep{fletcher16}. We compare the GRS spectrum with three models with reflecting layers at 500, 550, and 600~mbars. \textbf{The best fit to all \ammonia~ features has a reflecting layer at 570$\pm 30$~mbars where the \ammonia~ mole fraction is 40~ppm. If \ammonia~ is sub-saturated, the reflecting layer would be somewhat deeper (e.g. for 70\% relative humidity, the \ammonia~ mole fraction would be 40~ppm at 600~mbars)}. This is consistent with a cloud composed of ammonia ice.

\subsubsection{Middle cloud level}

Evidence for a cloud layer in the Great Red Spot between the ammonia and water clouds requires a combination of 5-\um~ and 7-\um~ data. Since the Earth's atmosphere is opaque near 7~\um, it is necessary to use spacecraft data. We used thermal infrared spectra from the CIRS investigation on the Cassini mission acquired as it flew by Jupiter en route to Saturn. As described in Section 2, we selected CIRS spectra from an NEB Hot Spot for comparison with the Great Red Spot. Of particular interest is a window to Jupiter's troposphere at 7.18~\um~(1392~\wn) due to a minimum in \methane~ opacity. This window was first exploited by \citet{matcheva05} to map Jupiter's belt-zone structure with CIRS data. The contribution function for a cloud-free atmosphere at 1392~\wn peaks near 1.3 bars. Thus, this spectral window sounds the atmosphere between the expected levels for ammonia clouds and ammonium hydrosulfide clouds on Jupiter. This allowed \citet{matcheva05} to study the variable opacity of \ammonia~ clouds on Jupiter without any confusion from deeper clouds.

A comparison of the radiance of the Great Red Spot to an NEB Hot Spot at 1392~\wn will give a rough estimate of the transmittance of the ammonia clouds within the GRS. In Fig. 11 we compare the GRS spectrum to that of the NEB Hot Spots. The radiance scale for the GRS in this spectral region is 0.25 times that of the NEB Hot Spot. The central peak at 1392~\wn is a continuum region bracketed by \methane~ absorption lines throughout the region between 1380 and 1435~\wnp. If the NEB Hot Spots are cloud-free for pressures less than 1.3~bars, then the transmittance of \ammonia~ clouds above the GRS would be 0.25 to match the radiance at 1392~\wnp.

\begin{figure}[!ht]
\begin{center}
\begin{tabular}{ll}
\hspace{-0.15in}
	 \includegraphics[width=3.0in]{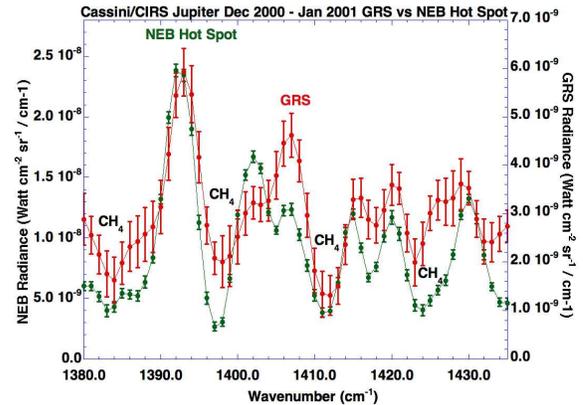}
\end{tabular}	\vspace{-0.1in}
   \caption {\footnotesize {Comparison between Cassini / CIRS spectra of the Great Red Spot and a Hot Spot in the North Equatorial Belt. The radiance scale for the GRS is 0.25 times that of the Hot Spot. These spectra have numerous \methane~ absorption lines. The continuum at 1392~\wn~ sounds the 1.3-bar level, well below the \ammonia~ clouds.}}\vspace{-0.2in}
   \label{fig11}
\end{center}  
\end{figure}

To provide indirect evidence for the existence of ammonium hydrosulfide clouds in the Great Red Spot, we combine information that we have learned from studying GRS spectra at 5~\um~ and at 7.18~\um.  We can use laboratory measurements of the imaginary index of refraction (k) of ammonia ice at 80~K by \citet{howett07a} to constrain the transmittance of the \ammonia~ cloud at 5~\um. \textbf{The value of k is $3.32\times 10^{-3}$ at 1400~\wn, $1.13\times 10^{-2}$ at 1900~\wn, and has values of $1.5\times 10^{-3}$ or less from 1950 to 2200~\wnp}. Thus, ammonia ice is less absorbing across the 5-\um~window than at 7.18~\um, except near the strong \ammonia~ gas (and ice) absorption band at 5.3~\um. Our model of the GRS requires a total cloud transmittance of 0.0185 at 4.66~\um~ above the line formation region which lies between 2 and 5~bars. From 5-\um~ data alone, this cloud could be entirely at the ammonia cloud level. By combining CIRS data at 7.18~\um~ with ground-based data at 5~\um, we suggest that the Great Red Spot contains clouds at both the expected levels for \ammonia~ and \amhs~ clouds. Adopting a value for the transmittance for the \ammonia~ cloud at 4.66~\um~ equal to that at 7.18~\um~ (0.25) would require an \amhs~ cloud transmittance of 0.074. Given that the \ammonia~ ice cloud will be less absorbing at 4.66~\um, the lower cloud must be the dominant absorber, with a transmittance less than or equal to 0.074. Thus, we conclude that the Great Red Spot \textbf{most likely} contains \amhs~ clouds that are more optically thick at 5~\um~ than the overlying \ammonia~ clouds.

\subsubsection{Summary of Cloud Model}

We present a schematic of our three-layer cloud model of the Great Red Spot in Fig.~12. The transmission of each cloud layer is shown next to a Galileo temperature, pressure profile. Thermal emission from the GRS originates at 5~bars from an opaque water cloud. It then traverses a thick \amhs~ cloud, followed by a thinner \ammonia~ cloud. \textbf{A second component of the GRS spectrum comes from sunlight reflecting off the \ammonia~ cloud with an albedo of 17$\pm3$\%. Fraunhofer lines provide evidence for this reflected component and their depth constrains the outgoing radiance to be 60$\pm10$\% \ reflected sunlight and 40$\mp10$\% \ thermal emission at 4.66~\um}.

\begin{figure}[!ht]
\begin{center}
\begin{tabular}{ll}
\hspace{-0.15in}
	 \includegraphics[width=3.0in]{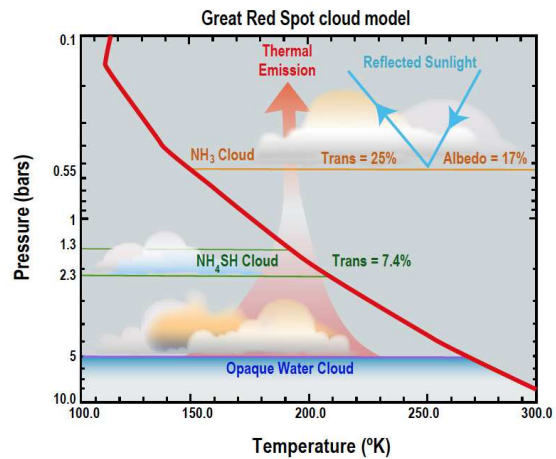}
\end{tabular}	\vspace{-0.1in}
   \caption {\footnotesize {Simplified cloud model for the Great Red Spot. The three cloud layers are shown \textbf{with significant opacity at 0.57$\pm0.03$ bar, $\sim$1.3 to 2.3~bars, and 5$\pm1$ bars} next to a Galileo Probe temperature, pressure profile.\textbf{The transmittances of these same three clouds are 25\%, 7.4\%, and 0\%, respectively. The upper cloud has an albedo of 17$\pm3$\%.}}}\vspace{-0.2in}
   \label{fig12}
\end{center}  
\end{figure}

\newpage
\subsection{Gas abundances}
\subsubsection{Water abundance}

We next modeled the Great Red Spot between 4.95 and 4.99~\um~ (2006-2020~\wnp) to constrain the water abundance. We used iSHELL data in this region rather than NIRSPEC due to the higher Doppler shift and spectral resolution. We used the cloud model described in the previous section and varied one model parameter. We assumed a saturated \water~ profile above an opaque cloud at 2, 5, and 7~bars.  Due to a Doppler shift of 28 km/sec jovian \water~ lines in the GRS are shifted by 0.19~\wn to the left of telluric water lines in Figures 13 and 14.

\begin{figure}[!ht]
\begin{center}
\begin{tabular}{ll}
\hspace{-0.15in}
	 \includegraphics[width=3.0in]{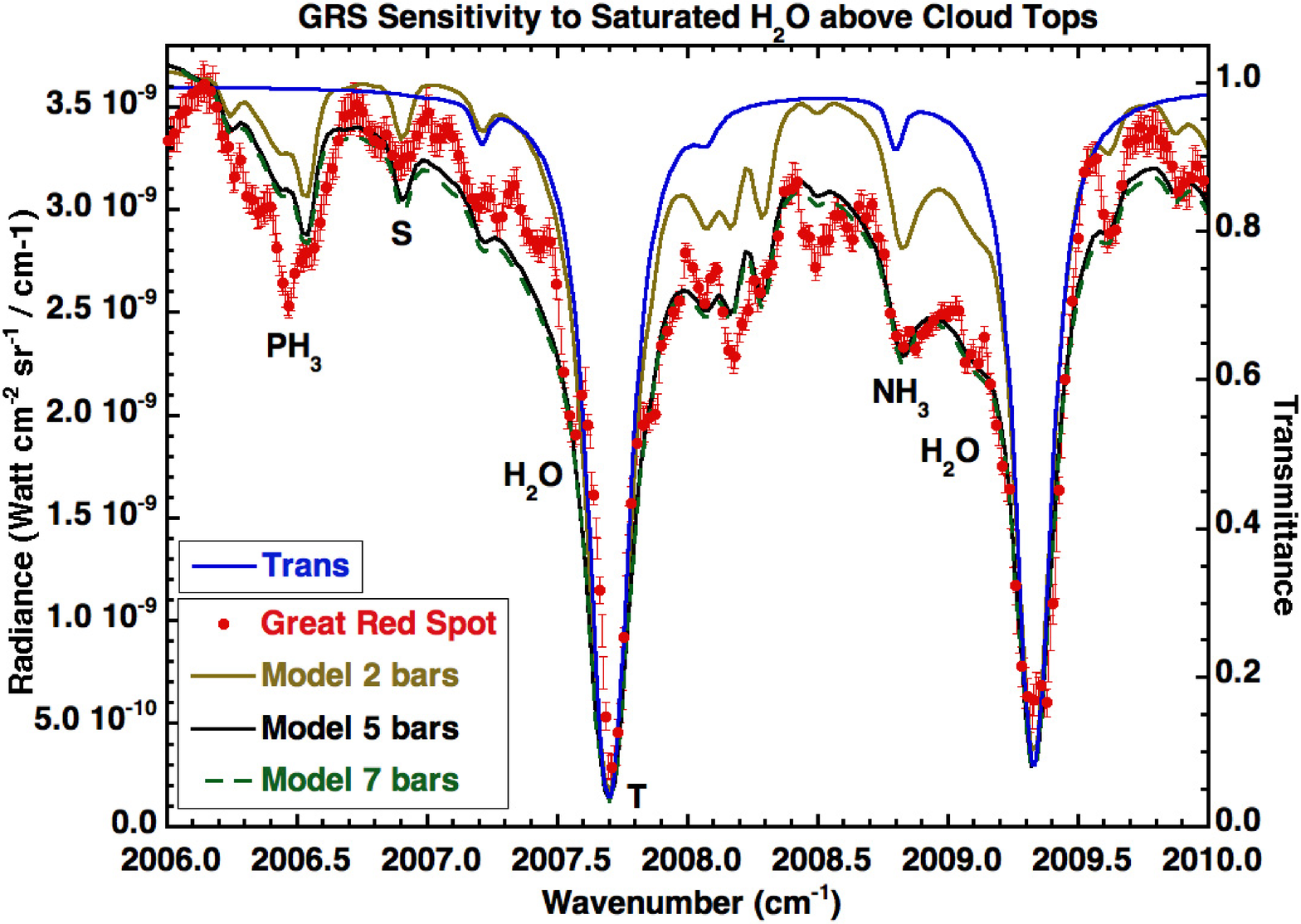}
\end{tabular}	\vspace{-0.1in}
   \caption {\footnotesize {Comparison between iSHELL spectrum of the Great Red Spot at 4.98~\um~ and three models that differ only in the pressure of the lower boundary, assumed to be an opaque cloud. A saturated \water~ profile is used for all 3 models. The model with an opaque cloud at 2 bars is excluded, while models with an opaque cloud at 5 and 7 bars fit the spectrum equally well.}}\vspace{-0.2in}
   \label{fig13}
\end{center}  
\end{figure}   

Note that the observed \water~ line profiles at  2007.5, 2009.1, 2016.4, and 2018.0 are broader than the telluric \water~features. The model with an opaque base at 2 bars (gold curve) fails to fit the spectrum because nearly all of the jovian \water~ is frozen out at these levels with temperatures less than 209~K. Models with lower boundaries at 5~bars (274~K) and 7~bars (304~K) fit the 4.96-\um~ spectrum equally well. The spectrum of the GRS at 4.67~\um~ (2138~\wnp) is also sensitive to \water. The spectrum shown in Fig.~8 is matched better by a model with an opaque cloud at 5~ bars, rather than 7~bars.  Thus, we adopt a model with a saturated \water~ distribution above an opaque water cloud at 5~bars.
 
\begin{figure}[!ht]
\begin{center}
\begin{tabular}{ll}
\hspace{-0.15in}
	 \includegraphics[width=3.0in]{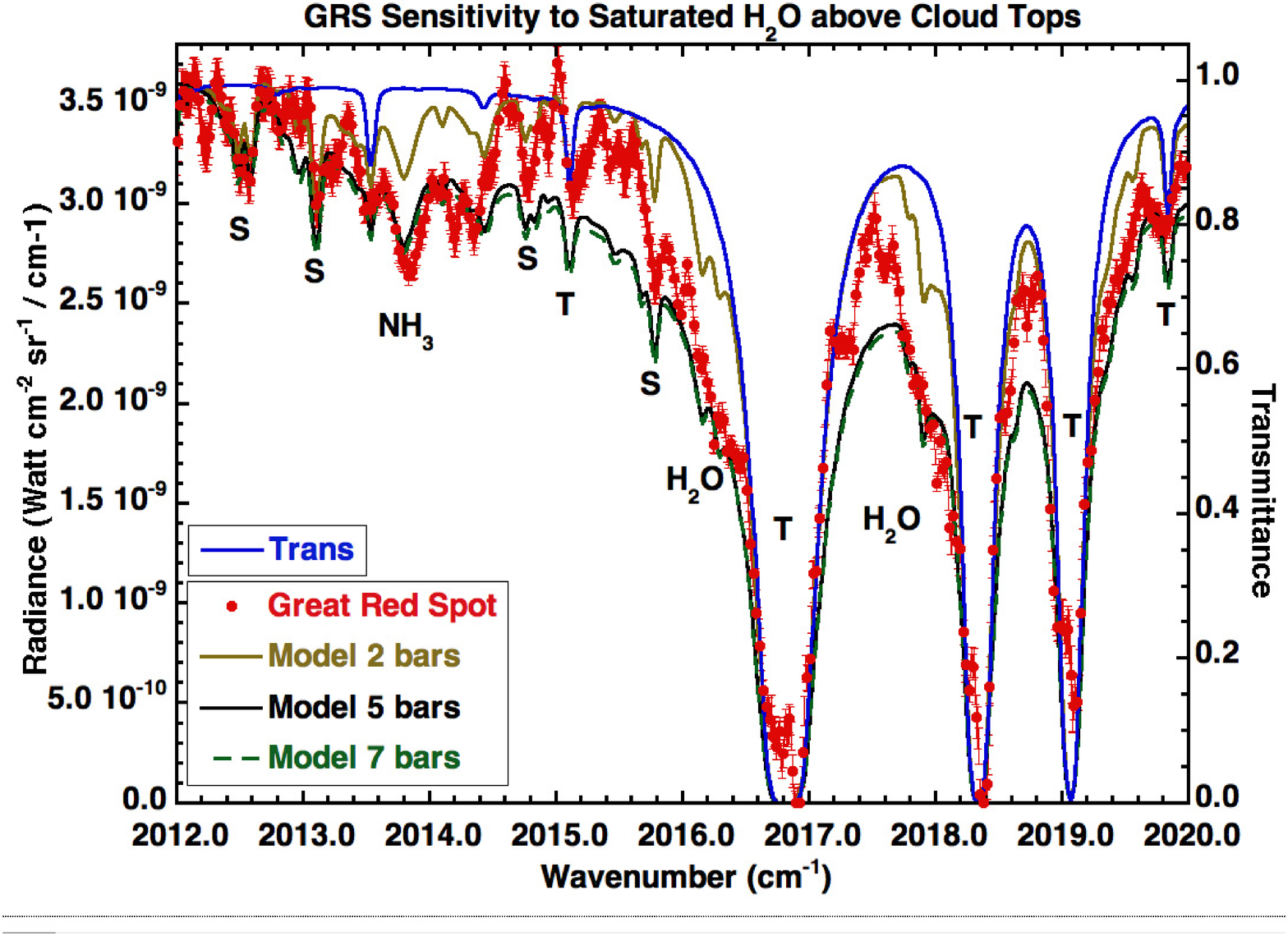}
\end{tabular}	\vspace{-0.1in}
   \caption {\footnotesize {Same as in Fig. 13, but for \water~ lines at 4.96~\um. Models with an opaque cloud at 5 and 7 bars fit the wings of \water~ lines at 2016.4 and 2018.0 \wn equally well, whereas the 2-bar model does not fit the spectrum.}}\vspace{-0.2in}
   \label{fig14}
\end{center}  
\end{figure} 

\subsubsection{Ammonia abundance}

We used the iSHELL spectrum of the Great Red Spot at 5.16~\um~ to constrain the deep abundance of ammonia. We present in Fig.~15 a portion of the GRS spectrum between 1937 and 1941~\wnp. The two strong absorption lines of \ammonia~ with rest frequencies of 1938.97 and 1939.56 belong to the \twonutwo~band. They sound the 3-bar level on Jupiter. These lines are Doppler-shifted by 0.18~\wn in the spectrum of the GRS. Using the cloud model described earlier, we calculated the spectrum for three abundances of \ammonia, namely, 100, 200, and 400~ppm. The model with 200~ppm provides the best fit to the spectrum. This same abundance fits two weaker \ammonia \ features at 2008.8~\wn (Fig.~13) and 2013.8~\wn (Fig.~14). \textbf{We adopt a value of 200$\pm 50$~ppm \ammonia \ in the Great Red Spot}.

\begin{figure}[!ht]
\begin{center}
\begin{tabular}{ll}
\hspace{-0.15in}
	 \includegraphics[width=3.0in]{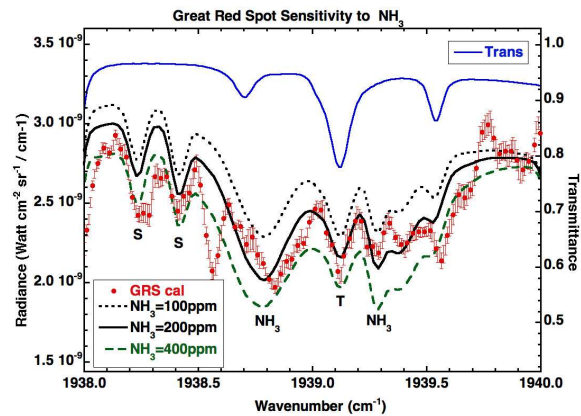}
\end{tabular}	\vspace{-0.1in}
   \caption {\footnotesize {Comparison between iSHELL spectrum of the Great Red Spot at 5.16~\um~ and three models with \ammonia~ mole fractions of 100, 200, and 400~ppm in the deep atmosphere. \textbf{We adopt a value of 200$\pm 50$~ppm \ammonia \ in the GRS.}}}\vspace{-0.2in}
   \label{fig15}
\end{center}  
\end{figure}   

In Fig.~16 we present the vertical profiles of \water~ and \ammonia~ that provide the best fit to our 5-\um~ spectra. Water vapor follows a saturated profile above the water cloud at 5~bars. Ammonia has a mole fraction of 200~ppm in the deep atmosphere and is assumed to follow a saturated profile for pressures less than 0.7~bars.

\begin{figure}[!ht]
\begin{center}
\begin{tabular}{ll}
\hspace{-0.12in}
	 \includegraphics[width=2.95in]{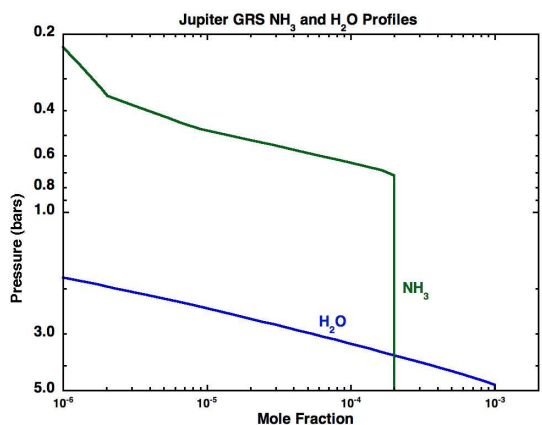}
\end{tabular}	\vspace{-0.1in}
   \caption {\footnotesize {Vertical profiles of \water~ and \ammonia~ in the Great Red Spot used to calculate synthetic spectra. The water mole fraction is 1000~ppm at 5~bars and it follows a saturated profile above this level. Ammonia is 200~ppm in the deep atmosphere. It begins to condense at 0.7~bars and it follows a saturated profile in the upper troposphere. }}\vspace{-0.2in}
   \label{fig16}
\end{center}  
\end{figure} 

\subsubsection{Phosphine abundance}

We used the iSHELL spectrum of the Great Red Spot at 4.66~\um~ to constrain the deep abundance of phosphine (\ph). We present in Fig.~17 a portion of the GRS spectrum between 2142 and 2145~\wnp. Using the cloud model described earlier, we calculated the spectrum for three abundances of \ph, namely, 0.4, 0.8, and 1.2~ppm for pressures greater than 0.7~bar. Our \ph~ profile falls off in the upper troposphere and is zero in the stratosphere in order to be consistent with previous studies (e.g. \citet{irwin04}, and \citet{fletcher10}); however, the 4.66-\um~ spectrum is not sensitive to pressures less than 0.7~bar. The model with 0.8~ppm provides the best fit to the spectrum. \textbf{We adopt a value of 0.8$\pm 0.2$~ppm \ph \ in the Great Red Spot}.

In Fig.~18 we modeled a portion of the warm collar surrounding the Great Red Spot located 3.3\arcsec~ west of its center.
In order to fit the broad \dm~ absorption lines, there must be no significant cloud opacity at the predicted level of the water cloud. We used the \water~ profile measured by the Galileo Probe for pressures greater than 4.5~bars. There are four \ph~ absorption lines in this portion of the spectrum. The best fit to the Warm Collar spectrum requires 0.4~ppm \ph, which is half the abundance required to fit the GRS. \textbf{We adopt a value of 0.4$\pm 0.1$~ppm \ph \ in the Warm Collar}.

\begin{figure}[!ht]
\begin{center}
\begin{tabular}{ll}
\hspace{-0.15in}
	 \includegraphics[width=3.0in]{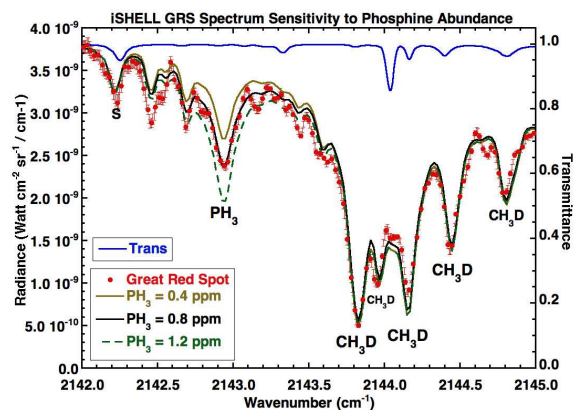}
\end{tabular}	\vspace{-0.1in}
   \caption {\footnotesize {Comparison between iSHELL spectrum of the Great Red Spot and three models with \ph~ mole fractions of 0.4, 0.8, and 1.2~ppm in the deep atmosphere. \textbf{We adopt a value of 0.8$\pm 0.2$~ppm \ph \ in the GRS.}}}\vspace{-0.2in}
   \label{fig17}
\end{center}  
\end{figure}   

\begin{figure}[!ht]
\begin{center}
\begin{tabular}{ll}
\hspace{-0.15in}
	 \includegraphics[width=3.0in]{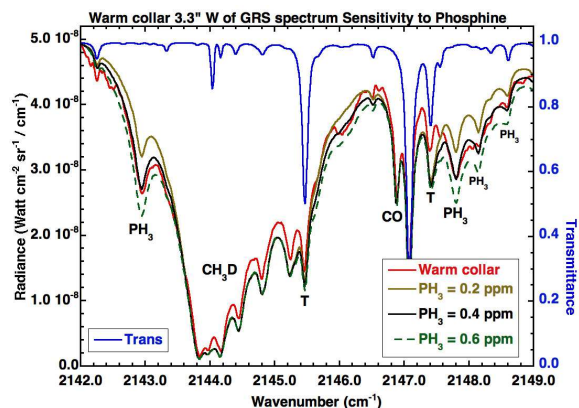}
\end{tabular}	\vspace{-0.1in}
   \caption {\footnotesize {Comparison between iSHELL spectrum of the Warm Collar and three models with \ph~ mole fractions of 0.2, 0.4, and 0.6~ppm in the deep atmosphere. \textbf{We adopt a value of 0.4$\pm 0.1$~ppm \ph \ in the Warm Collar.}}}\vspace{-0.2in}
   \label{fig18}
\end{center}  
\end{figure}

\vspace{0.2in}
\subsubsection{Germane abundance}

We used the iSHELL spectrum of the Great Red Spot at 4.66~\um~ to constrain the abundance of germane (\germane). We present in Fig.~19 and Fig.~20 portions of the GRS spectrum near 2139 and 2150~\wnp. We detect all 5 naturally occurring isotopes of germanium in \germane. Assuming terrestrial isotopic ratios, we calculated the spectrum for three abundances of \germane~: 0.3, 0.4, and 0.6~ppb. The model with 0.3~ppb fits an absorption feature of  \gefour~ at 2139~\wn, while 0.4~ppb is required to fit features near 2150~\wn. \textbf{We therefore adopt a value of 0.35$\pm0.05$~ppb for \germane~ in the Great Red Spot}.

\begin{figure}[!ht]
\begin{center}
\begin{tabular}{ll}
\hspace{-0.15in}
	 \includegraphics[width=3.0in]{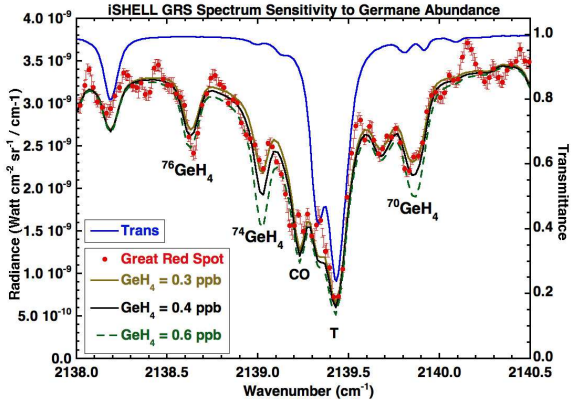}
\end{tabular}	\vspace{-0.1in}
   \caption {\footnotesize {Comparison between iSHELL spectrum of the Great Red Spot and three models with \germane~ mole fractions of 0.3, 0.4, and 0.6~ppb. Three isotopologues of \germane~ are shown. The model with 0.3~ppb \germane~ provides the best fit to this portion of the spectrum.}}\vspace{-0.2in}
   \label{fig19}
\end{center}  
\end{figure}   

\begin{figure}[!ht]
\begin{center}
\begin{tabular}{ll}
\hspace{-0.15in}
	 \includegraphics[width=3.0in]{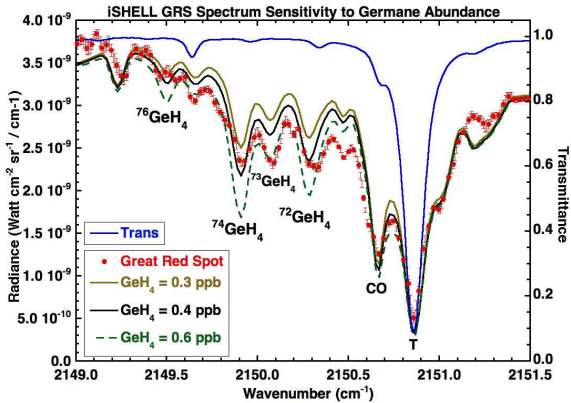}
\end{tabular}	\vspace{-0.1in}
   \caption {\footnotesize {Comparison between iSHELL spectrum of the Great Red Spot and three models with \germane~ mole fractions of 0.3, 0.4, and 0.6~ppb. Four isotopologues of \germane~ are shown. The model with 0.4~ppb \germane~ provides the best fit to this portion of the spectrum.}}\vspace{-0.2in}
   \label{fig20}
\end{center}  
\end{figure} 

\vspace{0.4in}
\subsubsection{Carbon monoxide abundance}

We used the iSHELL spectrum of the Great Red Spot at 4.66~\um~ to constrain the tropospheric abundance of carbon monoxide. We present in Fig.~21 and Fig.~22 two portions of the GRS spectrum near 2147 and 2151~\wnp. Here the combination of high Doppler shift (28 km/sec) and high spectral resolution (0.06~\wnp) permits a clean separation of 0.20~\wnp \ (more than 3 resolution elements) between the telluric CO line, marked T, and the jovian feature. Using the cloud model described earlier, we calculated the spectrum for three tropospheric abundances of CO: 0.4, 0.8, and 1.2~ppb. The model with 0.8~ppb in the deep atmosphere provides the best fit to the spectrum. \textbf{We adopt a value of 0.8$\pm 0.2$~ppb CO in the troposphere}. All models used the stratospheric profile for CO of \citet{bezard02}.

\begin{figure}[!ht]
\begin{center}
\begin{tabular}{ll}
\hspace{-0.15in}
	 \includegraphics[width=3.0in]{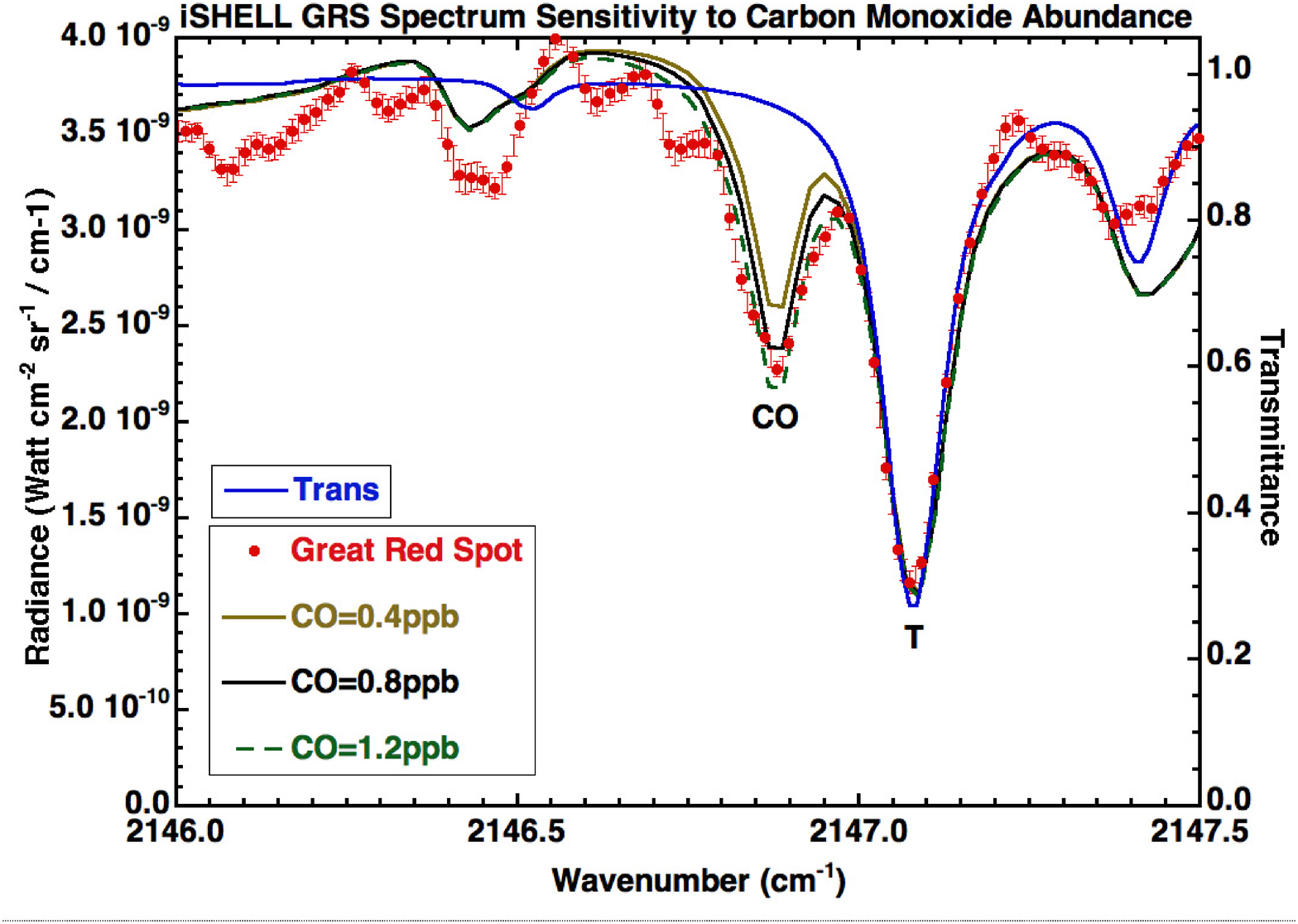}
\end{tabular}	\vspace{-0.1in}
   \caption {\footnotesize {Comparison between iSHELL spectrum of the Great Red Spot and three models with CO mole fractions of 0.4, 0.8, and 1.2~ppb in the troposphere. The R0 line of CO on Jupiter is well separated from its telluric counterpart (T). \textbf{We adopt a value of 0.8$\pm 0.2$~ppb CO in the troposphere.}}}\vspace{-0.2in}
   \label{fig21}
\end{center}  
\end{figure}

\begin{figure}[!ht]
\begin{center}
\begin{tabular}{ll}
\hspace{-0.15in}
	 \includegraphics[width=3.0in]{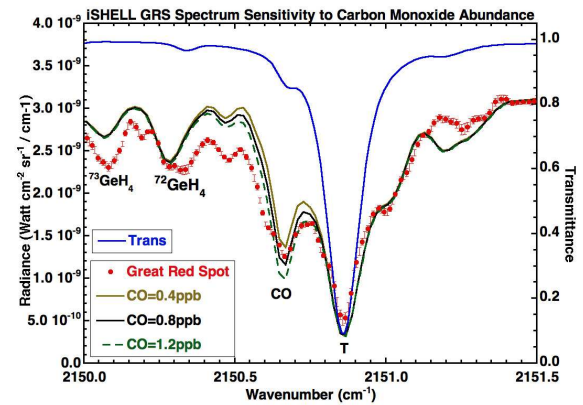}
\end{tabular}	\vspace{-0.1in}
   \caption {\footnotesize {Same as Fig.~21 for the R1 line of CO. \textbf{We adopt a value of 0.8$\pm 0.2$~ppb CO in the troposphere.}}}\vspace{-0.2in}
   \label{fig22}
\end{center}  
\end{figure}





\section{DISCUSSION}
\subsection{Cloud Structure}

Using a simplified 3-cloud model, we obtained a good fit to the 5-\um~ spectrum of the Great Red Spot using a framework that is consistent with thermochemical models. Future studies can build on and test this framework by adding cloud microphysics, which would include modeling the vertical extent of each cloud layer. In addition, improved models would include wavelength-dependent absorption coefficients of \ammonia~ ice and \amhs~ ice across the 5-\um~ window, and incorporate scattering.

\textbf{Our discovery of an opaque cloud at 5$\pm1$~bars (almost certainly a water cloud)} in the Great Red
Spot may suggest that the vortex extends much deeper than the
water cloud layer. \citet{marcus13} developed a model of
secondary circulation within jovian anticyclones that maintains
their key thermodynamic features: a high-pressure core with its
associated geostrophic winds, sandwiched between a cold,
high-density lid at the top of the vortex, and a warm,
low-density anomaly at the base of the vortex. The cold lid in
this scenario is a well-known observational feature \citep{cheng08, fletcher10}, and \citet{wong11} found that static stability measurements and models agreed with vortex
models at the upper tropospheric levels where the cold lid is
seen. \citet{wong11} also found that the low-density anomaly
at the base of the vortex is consistent with stratification
produced by condensation of a supersolar water cloud. If the GRS indeed had its base in the water
cloud layer, then the warm, low-density anomaly there should
generally inhibit water cloud formation, a scenario that is
contradicted by our results in Figures 7 and 8. In fact, if
the \citet{marcus13} model holds, our observation of 5-bar
water clouds in the GRS may suggest that the vortex midplane lies
below the 5-bar level, so that the water cloud layer is within
the upper half of the vortex, where the secondary circulation is
dominated by upwelling. Such a large vertical extent would also
be consistent with preliminary Juno Microwave Radiometer
observations of the GRS \citep{li17b}, as well as dynamical
simulations showing that the vortex can extend well into the free
convective zone \citep{chan13}.

\textbf{Our derived pressure of the reflecting layer at 570$\pm30$~mbars in the GRS} is consistent with previous studies of upper clouds by \citet{banfield98}, but it is inconsistent with a layer at 900~mbars as suggested by \citet{simon01}. Our evidence for a thick middle cloud is consistent with near-infrared spectra from Galileo NIMS. \citet{irwin01} investigated the question of whether the \ammonia~ cloud or a deeper cloud is responsible for the observed variability in the brightness of Jupiter at 5~\um. They studied Galileo NIMS data and found an anti-correlation between the brightness of Jupiter at 5~\um~ and that at 1.58~\um. Irwin et al. used a reflecting layer model to fit NIMS spectra from 0.7 to 2.5~\um~ of belts (bright at 5~\um, dark at 1.58~\um) and zones (dark at 5~\um, bright at 1.58~\um). Successful models required variations in cloud opacity to lie deeper than 1 bar. Their model of the Great Red Spot from NIMS data required increased \ammonia~ cloud opacity to fit 0.7 to 1.0~\um~ as well as increased cloud opacity between 1 and 2 bars to fit the spectrum from 1.0 to 2.5~\um. Thus, Irwin et al. provided evidence for two cloud layers in the GRS, but they could not determine the optical depths of each cloud separately.

The cloud structure derived for the GRS in this study is fairly similar to that derived for a cloudy feature in the South Tropical Zone (STZ) by \citet{bjoraker15}. In that paper we argued that mass flux into the upper cloud layers of the STZ is not dominated by horizontal transport, as suggested by  \citet{showman05}, but instead is driven by vertical transport from below. Jupiter's circulation in zones (and now in the Great Red Spot) therefore maintains the same sign of upwelling/downwelling across the full 0.5-5~bar weather layer. The same sign of upwelling/downwelling over this large an extent was also derived by \citet{depater10} from 5-micron bright rings around vortices. These authors suggested that vortices must extend vertically from at least the 4 to 7-bar level up to the tropopause. This seems to be the case for the Great Red Spot as well.

\subsection{Water abundance and the O/H ratio}

This study constitutes the first detection of gaseous water in Jupiter's Great Red Spot. Water follows a saturated profile in the GRS, similar to that of a cloudy feature in the STZ, but unlike the highly depleted \water~ profile observed in Hot Spots \citep{bjoraker15}. The pressure that we derived for the opaque lower cloud is useful for understanding the vertical extent and dynamics of the GRS. In addition, it provides a constraint on the deep water abundance of Jupiter as a whole, and therefore its O/H ratio, assuming that the lower cloud is, in fact, composed of water. The base of the water cloud is sensitive to the deep abundance of water because higher abundances lead to condensation at deeper levels. Our data provide the level of the cloud top, not the cloud base. Since the top is at a higher altitude than the base, cloud top constraints provide lower limits to the pressure of the cloud base, or, lower limits to the deep abundance of water. The GRS spectrum requires a water cloud top at $P\ge5$~bar. The cloud base is therefore found at $P > 5$~bars. Following Fig.~1 in \citet{wong08}, a cloud base at 5 bar corresponds to an O/H ratio of 1.1$\times$ solar (corrected to the new solar O/H ratio of Asplund et al. 2009). The O/H lower limit of 1.1$\times$ solar is consistent with, and more constraining than, the Galileo Probe lower limit of 0.48$\pm$0.16$\times$ solar \citep{wong04}. Our observations support either solar or super-solar enrichments of water in Jupiter, providing somewhat better constraints to planetary formation models (\citet{gautier01}, \citet{hersant04}, and \citet{wong08}).

\subsection{Ammonia abundance}

\textbf{Our retrieved \ammonia~ abundance of 200$\pm50$~ppm in the Great Red Spot} for pressures greater than 0.7~bars may be compared with other 5-\um~ studies. \citet{giles17b} obtained 5-\um~ spectra of Jupiter using the CRIRES instrument on the Very Large Telescope at the European Southern Observatory in Chile. They did not observe the Great Red Spot, but they did report the latitudinal variation of \ammonia~ acquired using spectra aligned north-south on Jupiter's central meridian. They found NH$_3$ mixing ratios less than 200 ppm at 1.6 bars for all latitudes using a strong NH$_3$ absorption feature (at 1939~cm$^{-1}$; see Fig. 15), and larger values for NH$_3$ at 3.3 bars using a weaker feature (at 1929 cm$^{-1}$). Some differences between this group's work and our results may be due to the very
different treatments of cloud opacity in the two studies. \citet{giles17b} initially used a single compact cloud layer at 0.8~bars. Later, they examined the effect of an additional cloud at 5~bars, following \citet{bjoraker15}. They also did not include sunlight reflected from an upper cloud layer.  \citet{giles17b} obtained 150~ppm \ammonia~ at 1.6~bars increasing to 250~ppm at 3.3~bars for a region of Jupiter at 20\deg~ South latitude on Jupiter, which is the latitude of the GRS. This is in good agreement with our results. However, this agreement may be fortuitous if \ammonia~ varies with longitude due to local dynamics.

Using microwave observations of Jupiter from the Very Large Array radio telescope, \citet{depater16} obtained longitude-resolved maps with sufficient spatial resolution (1200~km) to resolve the Great Red Spot. Using microwave channels from 5.49 to 17.38 GHz (1.7 to 5.5~cm), they obtained a vertical profile for \ammonia~ at the center of the GRS of 200~ppm at pressures less than 1.5~bars increasing to 570~ppm at deeper levels. Contribution functions for these wavelengths cover roughly 0.6 to 6~bars in Jupiter's troposphere, depending on the exact \ammonia~ profile (see Fig.~1). In a future paper we plan to model the GRS using an updated (and recalibrated) microwave spectrum between 5 and 25 GHz. We plan to include microwave opacity by clouds in addition to absorption by \ammonia, \hs, and \water~ (e.g., \citet{depater93, depater05}) for comparison with the current study at 5~\um.

Our deep (0.7 - 5 bars) \ammonia~ concentration of 200~ppm is a low value, and has implications for the 
vertical profile of \ammonia~ in Jupiter's atmosphere both inside and outside of the GRS. Assuming that the 
internal circulation of the GRS is somewhat isolated from its exterior, the 200~ppm concentration of 
\ammonia~ is likely to persist all throughout the interior of the vortex, even below the water cloud 
that limits our observational sensitivity to pressures less than 5~bars. Our deep GRS value is lower than the 
concentration of about 400~ppm in the Galileo Probe entry site near 5 bars \citep{hanley09}, lower than the 
concentration of 570~ppm at the deepest levels of the Galileo Probe entry site \citep{wong04}, and lower 
than the deepest concentration of 360~ppm derived from Juno MWR data \citep{li17}.

Measurements of the \ammonia~ profile in the Galileo Probe Entry Site, a 5-\um~ Hot Spot, found a gradually 
increasing concentration of \ammonia, down to a well-mixed level near 8~bars, coincident with a layer 
of higher static stability \citep{magalhaes02}.  The 8-bar level possibly represented the ammonia cloud 
base \citep{wong09} after having been deflected downward by the Rossby wave system responsible for 
creating these Hot Spots \citep{showman98, friedson05}. This scenario implied that Jupiter's \ammonia~ 
concentration should be around 570~ppm everywhere below the cloud base, a high value that was difficult 
to reconcile with disk-averaged microwave spectra \citep{depater01}. But spatially resolved microwave 
spectroscopy can explain these discrepancies by showing that high (570~ppm) ammonia below the cloud base 
indeed occurs in plumes located near (and to the south of) 5-\um~ Hot Spots, while other regions of the planet 
are less ammonia-rich \citep{depater16}. This is qualitatively consistent with the Juno MWR findings of 
depleted (200~ppm) ammonia in the 3 - 10~bar range, over a wide range of latitudes [10 - 40$^{\circ}$ 
S and 20 - 40$^{\circ}$ N,  (see \citet{li17}). Although \ammonia~ within the GRS should be vertically mixed by the internal circulation of the vortex, the composition of the air within the GRS should generally reflect the composition of its 
surroundings.

\subsection{Phosphine abundance}

Phosphine is a disequilibrium species in Jupiter's atmosphere. Equilibrium models by \citet{fegley94} show that \ph~ should be abundant at 1500~K in the deep atmosphere, but it would be converted to P$_{4}$O$_{6}$ at colder levels. The presence of \ph~ at the 5-bar level requires vertical transport from great depth on a time scale faster than the chemical conversion to P$_{4}$O$_{6}$ with an eddy diffusion coefficient (K) on the order of 10$^{8}$ cm$^{2}$ sec$^{-1}$. The value of K was constrained by measurements of \ph~ primarily in Hot Spots (e.g. \citet{bjoraker86b}). Our new measurements of \ph \ in the GRS indicate that vertical transport rates in the GRS may be higher than those in Hot Spots. \textbf{An alternate disequilibrium model was developed by \citet{wang16}. In this model \ph \ is converted to H$_{3}$PO$_{4}$ instead of P$_{4}$O$_{6}$. They found that \ph \ is relatively insensitive to K, and thus should not vary with location on Jupiter.} 

\textbf{Our \ph~ abundance in the GRS at the 5-bar level (0.8$\pm0.2$~ppm) is a factor of 2 higher than in the warm collar surrounding the GRS (0.4$\pm0.1$~ppm)}. \citet{bjoraker15} measured 0.45~ppm \ph~ in an SEB Hot Spot at 17\deg S and 0.7~ppm \ph~ in a region in the STZ at 32\deg S. These four measurements provide evidence for an enhancement in \ph \ at the 5-bar level in cloudy regions (GRS and STZ) compared with regions lacking an opaque water cloud (an SEB Hot Spot and the Warm Collar). This may be a consequence of enhanced convection in cloudy regions on Jupiter. \textbf{These measurements are not consistent with the predictions of the model developed by \citet{wang16}.}

\citet{fletcher16} reported 1.2~ppm \ph~ over the Great Red Spot at 0.5~bars using IRTF/TEXES data at 10~\um. The \ph~ abundance fell off to 0.4 to 0.6~ppm in the warm collar surrounding the GRS. This falloff from the center of the GRS to the warm collar is in qualitative agreement with our measurements. However, our 0.8-ppm value for \ph~ at 5~bars is lower than the value at 0.5~bars (1.2~ppm) obtained using absorption features at 10~\um. \citet{fletcher09}  first noticed the discrepancy between \ph~retrievals at 5 and 10~\um. A larger abundance of \ph~ at 0.5 bars than at 5~bars would be inconsistent with the disequilibrium model. These authors attributed this discrepancy to possible errors in the line strengths of \ph. \textbf{This discrepancy motivated a laboratory study of \ph \ line strengths at 5~\um \ \citep{malathy14}. The newer measurements of \ph \ line intensities are about 7\% \ higher than the older lab data of \citet{tarrago92} that are on the GEISA line atlas \citep{husson05} that we used in this study. This is not sufficient to explain the 50\% difference between \ph \ mole fractions derived from 5 and 10~\um \ data.} Thus, this problem has yet to be resolved.

\subsection{Germane abundance}

Germane is also a disequilibrium species in Jupiter's atmosphere. Models by \citet{fegley94} \textbf{and by \citet{wang16}} show that \germane~ should be abundant at 2000~K in the deep atmosphere, but it would be converted to GeS at colder levels. The presence of \germane~ at the 5-bar level requires vertical transport from great depth with an eddy diffusion coefficient similar to that obtained for \ph, $\approx$ 10$^{8}$ cm$^{2}$ sec$^{-1}$. \textbf{Our retrieved value of 0.35$\pm0.05$~ppb in the GRS} is compatible with this vertical transport model.

\citet{giles17a} observed Jupiter's South Equatorial Belt (SEB) at 5~\um~ using the CRIRES instrument, as described above. They detected the strong Q-branch of the \nuthree~ band of \germane~ as well as the R3, R6, and R7 features. The R6 feature at 4.65~\um~ (2150.5~\wn) is the same feature that we observed in the Great Red Spot (see Fig.~20). Since the SEB does not have an opaque water cloud (see \citet{bjoraker15}) the pressure in the line formation region at 4.65~\um~ is higher than in the GRS. Therefore, absorption features due to various isotopologues of \germane~ are blended together in the SEB due to opacity broadening and pressure broadening by \h. By measuring \germane~ in a cloudy region such as the GRS, we can spectrally separate each individual isotopologue. In Fig.~19 and Fig.~20 we show absorption features of all 5 isotopic variants in the GRS. These are  \gezero, \getwo, \gethree, \gefour, and \gesix. The terrestrial relative abundances of germanium isotopes are 20.6\%, 27.5\%, 7.8\%, 36.5\%, and 7.7\% respectively \citep{berglund11}. We obtained line lists for all isotopologues of \germane~ from R. Giles (private communication, see \citet{giles17a} for details).

Using the same CRIRES dataset, \citet{giles17a} measured the latitudinal variation of the strong Q-branch of \germane~ on Jupiter. As with \ammonia, they found that it was difficult to separate the effects of spatially varying \germane~ from variations in the deep cloud structure. They found that they could fix \germane~ to 0.58~ppb, the value in the SEB, and allow the deep cloud to vary. Alternatively, they could allow \germane~ to vary with latitude, or they could permit both to vary. Their best fit model required both gas abundances and the deep cloud to vary with \germane~mole fractions ranging from 0.25 to 0.7~ppb. They obtained 0.45~ppb at 20\deg~ South, in good agreement with our results for the GRS.

\subsection{Carbon monoxide abundance}

There are two sources for CO on Jupiter. An external source of oxygen from meteoroids supplemented by large impacts such as from comet Shoemaker-Levy 9 results in the production of CO in Jupiter's stratosphere \citep{bezard02}. The second source of CO comes from \methane~ and \water~ in the deep atmosphere. Methane is the principal reservoir for carbon in Jupiter's reducing atmosphere, but at temperatures greater than 1000~K there is expected to be at least 1 ppb CO in chemical equilibrium with other carbon species including CH$_4$. Methane is the only equilibrium carbon species at the colder temperatures (273 K) probed at 5 $\mu$m. The unexpected detection of CO on Jupiter at 5~\um~ by \citet{beer75} led to the development of a disequilibrium model in which CO is transported from depth faster than the time scale to convert it back to \methane~ \citep{prinn77}.

The Prinn and Barshay model was updated by \citet{bezard02} using improved reaction rates. The abundance of CO at 5~bars is proportional to the deep O/H ratio and depends on the eddy diffusion coefficient (K). Thus, the tropospheric abundance of CO can provide  constraints on the deep water abundance as well as vertical transport rates. \citet{bezard02} measured 1.0 $\pm$ 0.2 ppb CO at 6~bars in an NEB Hot Spot.

\textbf{Our abundance of 0.8$\pm0.2$~ppb CO in the GRS} is the first measurement of CO in a region of Jupiter with thick clouds. Our CO results, when combined with our measurements of \ph~ and \germane, suggest that vertical transport rates in the GRS are similar to or somewhat higher than those in Hot Spots. A value of 10$^{8}$ cm$^{2}$ sec$^{-1}$ for the eddy diffusion coefficient and 0.8~ppb CO would correspond to an O/H ratio of 4 times solar in the Bezard et al. (2002) model using the more recent solar abundances of \citet{asplund09}. \citet{bezard02} estimated that K lies in a range between 4x10$^{7}$ and 4x10$^{9}$ cm$^{2}$ sec$^{-1}$. With these large uncertainties in K, \citet{bezard02} were only able to constrain Jupiter's O/H ratio to be between 0.35 and 15 times solar (after correction to the solar value of \citet{asplund09}). Using our lower limit to O/H of 1.1 times solar derived from the pressure of the water cloud, we can narrow the range of the O/H ratio to be between 1.1 and 12 times solar.

\textbf{\citet{wang15} proposed a new formulation for Jupiter's eddy diffusion coefficient. They investigated the range of water enrichment required to fit Jupiter's CO abundance using two different CO kinetic models. Using kinetics from \citet{visscher11}, and a CO abundance of 1~ppb from \citet{bezard02}, they inferred O/H enrichments ranging between 0.1 and 0.75 times solar. Using an alternate chemical kinetics model originally applied to hot Jupiters \citep{venot12}, they constrained the O/H ratio to be between 3 and 11 times solar. Our lower limit to O/H of 1.1 times solar derived from the pressure of the water cloud is consistent with the \citet{venot12} kinetic model, but not with the model of \citet{visscher11}. Using our CO value of 0.8~ppb in the Great Red Spot and the \citet{venot12} kinetic model, we obtain O/H enrichments between 2.4 and 9 times solar}.



\acknowledgments

The data presented here were obtained using NASA's Infrared Telescope Facility as well as using the Keck telescope. The W. M. Keck Observatory is operated as a scientific partnership among the California Institute of Technology, the University of California, and the National Aeronautics and Space Administration.  The Observatory was made possible by the generous financial support of the W. M. Keck Foundation. The authors extend special thanks to those of Hawaiian ancestry on whose sacred mountain we are privileged to be guests. Without their generous hospitality, none of the observations presented would have been possible. This research was supported by the NASA Planetary Astronomy grant NNX14AJ43G to UC-Berkeley and NASA Solar System Observations grant NNX15AJ41G to NASA/Goddard and UC-Berkeley. Context maps used observations made with the NASA/ESA Hubble Space Telescope, obtained from the Data Archive at the Space Telescope Science Institute, 
which is operated by the Association of Universities for Research in Astronomy, Inc., under NASA contract NAS 5-26555. These observations are associated with programs GO-13067 and GO-14756. 



\facilities{ IRTF (iSHELL, SpeX, NSFCam, CSHELL), Keck:II (NIRSPEC), Cassini (CIRS, ISS), Voyager (IRIS), HST (WFC3), Galilep Probe (ASI)}

\bibliography{Jupiter_Saturn.bib}
\bibliographystyle{aasjournal}


\clearpage



\end{document}